\documentclass[pra,showpacs,array]{revtex4}
\usepackage{graphicx}
\usepackage{dcolumn} 
\usepackage{bm}      

\newcommand{\dket}[1]{ \left| #1 \right\rangle }
\newcommand{\dbra}[1]{ \left\langle #1 \right| }

\newcommand{\ba}[1]{\begin{array}{#1} }
\newcommand{\ea}{\end{array}}
\newcommand{\rt}{ \frac{1}{\sqrt{2}} }
\newcommand{\ceqn}[1]{(\ref{#1})}
\newcommand{\cfig}[1]{Figure \ref{#1}}

\begin{document}

\title{Ground State Entanglement in a Combination of Star And Ring Geometries Of Interacting Spins}
\author{A. Hutton$^1$, S. Bose$^2$}
\affiliation{$^1$Centre for Quantum Computation, Clarendon
Laboratory, University of Oxford, Oxford OX1 3PU, UK\\
$^2$Department of Physics and Astronomy, University College
London, Gower St., London WC1E 6BT, UK}

\begin{abstract}
We compare a star and a ring network of interacting spins in terms
of the entanglement they can provide between the nearest and the
next to nearest neighbor spins in the ground state. We then
investigate whether this entanglement can be optimized by allowing
the system to interact through a weighted combination of the star
and the ring geometries. We find that such a weighted combination
is indeed optimal in certain circumstances for providing the
highest entanglement between two chosen spins. The entanglement
shows jumps and counterintuitive behavior as the relative
weighting of the star and the ring interactions is varied. We give
an exact mathematical explanation of the behavior for a five qubit
system (four spins in a ring and a central spin) and an intuitive
explanation for larger systems. For the case of four spins in a
ring plus a central spin, we demonstrate how a four qubit GHZ
state can be generated as a simple derivative of the ground state.
Our calculations also demonstrate that some of the multi-particle
entangled states derivable from the ground state of a star network
are sufficiently robust to the presence of nearest neighbor ring
interactions.
\end{abstract}

\maketitle

\section{Introduction}


The entanglement present in natural spin systems
\cite{EntRings,thermentA,ThermE-1D,EntXY,ThermXXZ,wang01-3,wangxx-2,IsingVarB,kamta,falci-1,falci-2,XXRings,ibose,arul,plenio,guifre,vladimir}
has been a subject of serious interest in recent years. It is
believed that this entanglement can even have consequences on the
macroscopic properties of such systems
\cite{ThermE-1D,aeppli-vedral-1,aeppli-vedral-2,vedral2}. The
entanglement is found to exhibit interesting behavior near the
points of quantum phase transitions
\cite{sachdev,ThermE-1D,IsingVarB,falci-1,falci-2,ibose,guifre,vladimir}.
Ground states of some finite systems can serve as a convenient
template for generating multipartite entangled states
\cite{starspin}. Creating the state
requires only cooling the system down. 

A number of spin structures can be investigated to determine their
entanglement properties in the ground state, in particular a ring
\cite{EntRings,XXRings}, as well as other lattice structures
\cite{ibose}. However, 1D chains and lattices of various
dimensions are not the {\em only} physical systems whose
fabrication is possible with current technology.  It is possible
to extend the above line of research on entanglement in spin
systems to other than spin chains. In particular, various
technologies have evolved which can make any member of an array of
qubits interact with any other member
\cite{ImplMTraps,ImplSiNuc,JJunctQubits,QD+CavQED,ImplCavQED}. One
such structure is the star in which a number of spins interact
through one central spin \cite{starspin}. Apart from our work on
the ground state entanglement in a spin-star \cite{starspin},
recently interesting non-Markovian dynamics \cite{burgarth},
quantum cloning \cite{dechiara} and quantum gates \cite{benjamin3}
in such a system has been investigated by other authors.

\begin{figure}
\begin{tabular}{ccc}
\includegraphics[width=1.85in, clip]{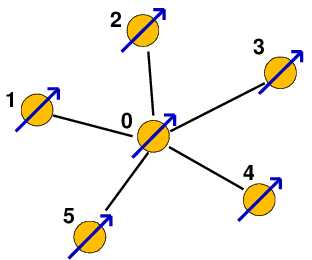} &
\includegraphics[width=1.85in, clip]{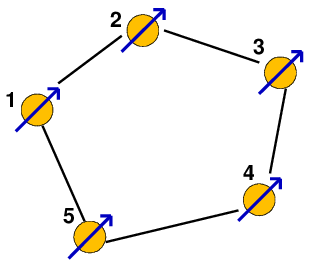} &
\includegraphics[width=1.85in, clip]{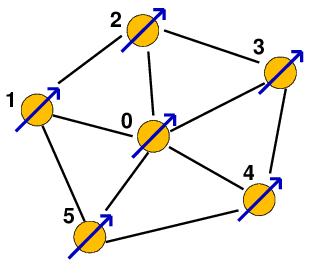} \\
(a) & (b) & (c) \\
\end{tabular}
\caption{The star (a), ring (b) and  star-ring combination (c)
spin models illustrated with $N=5$ spins placed in a ring with one
spin at the center. The lines connect interacting pairs of spins.}
\label{fig:spinpiccy}
\end{figure}

In this paper, we first briefly compare the star model and the
ring model for spins interacting via the $XX$ interaction, which
is physically realizable \cite{QD+CavQED,benjamin2}, to find out
whether one model or the other is better for establishing
entanglement between two spins (in the ground state). Then, as the
main focus of the paper, we study entanglement in a model in which
the spins interact simultaneously through both means. We study
both nearest and next to nearest neighbour entanglement between
spins. The system where the spins can interact both with their
neighbours and a central spin (a combination of the ring and star)
also represents a logical continuation of the work on entangled
rings \cite{EntRings,XXRings}, especially that of Wang
\cite{XXRings}. It simply considers the consequences of adding a
central spin to a ring of spins, so that the spins can also
interact via this central spin. Note that this is different to a
star network in which spins interact \emph{only} with a central
spin. In addition, it is also a natural extension of our earlier
work on a star network of spins \cite{starspin}. In practice, in a
spin star, unwanted interactions between the outer spins would
exist due to their physical proximity and thus our current
investigation considers how much these unwanted interactions would
modify the properties of the unpolluted star system (such as the
ability of the star system to produce interesting multi-particle
entangled states as the ground state).

    Our work is also motivated by the fact that
entanglement can show interesting behavior at points of quantum
phase transitions \cite{sachdev}, such as scaling
\cite{falci-1,falci-2,guifre,vladimir} and macroscopic jumps
\cite{ibose}. Very often frustration due to competing
non-commuting terms in a Hamiltonian is the cause of the curious
behavior of entanglement at a quantum phase transition
\cite{ibose,dawson}. By combining a star system Hamiltonian and a
ring system Hamiltonian, we precisely intend to create such a
frustration between competing ordering tendencies. The system of
our current paper is finite, and the two parts of the Hamiltonian,
namely the star system part and the ring system part, do not
commute, so we do not expect a quantum phase transition in our
system \cite{sachdev}. Nonetheless, as we will show, the
competition between two different parts of our Hamiltonian leads
to sharp changes (``jumps" in the same sense as Ref.\cite{ibose})
in the entanglement as the relative strength of the two terms is
varied. In addition to sharp changes, we will also show that the
magnitude of entanglement changes in a counterintuitive manner as
the relative strength of the two terms is varied. We will provide
a heuristic explanation for the observed behavior of entanglement.
Moreover, we will explicitly solve a system of four spins in a
ring interacting with a common central spin and show that this
system can be used to produce a Greenberger-Horne-Zeilinger (GHZ)
state \cite{ghz} as a simple derivative of the ground state, a
feat not achieved yet, to our knowledge, by any other spin
Hamiltonian.

 \cfig{fig:spinpiccy} depicts schematically the star and ring spin
models.  For both models the qubits in the outer ring will be
referred to as the `outer qubits'.  \cfig{fig:spinpiccy}(a)
depicts the star model, in which the outer qubits interact only
with a central qubit. \cfig{fig:spinpiccy}(b) depicts the ring
model, in which the outer qubits interact with their nearest
neighbours in the ring, while \cfig{fig:spinpiccy}(c) illustrates
a model where qubits interact both with their nearest neighbours
and with a central spin.  The outer qubits are labelled $1$ to
$N$, while the central qubit in the star model is labelled $0$.
The Hamiltonians for the two models are

\[ \ba{l}
H_{\text{ring}} = \sum_{i=1}^{N} \left( \sigma_{x}^{i} \sigma_{x}^{i+1} + \sigma_{y}^{i} \sigma_{y}^{i+1} \right) \\
H_{\text{star}} = \sum_{i=1}^{N} \left( \sigma_{x}^{0} \sigma_{x}^{i} + \sigma_{y}^{0} \sigma_{y}^{i} \right)
\ea \]
with periodic boundary conditions i.e.\ $N+1 = 1$.

The measure we will use for entanglement between two qubits is the entanglement of formation \cite{BigMixed, EntForm}.  Specifically, we will use the concurrence \cite{EntForm0,EntForm}, of which the entanglement of formation is a monotonic function.  To determine the concurrence between two qubits, we firstly trace out the other qubits in the model and then calculate the concurrence of the remaining two.

The two models have in common the fact that the Hamiltonian commutes with the total spin in the $z$ direction i.e.\
\[ \ba{l}
\left[ H_{\text{ring}}, \sum_{i=1}^N \sigma_z^{i} \right] = 0 \\
\left[ H_{\text{star}}, \sum_{i=0}^N \sigma_z^{i} \right] = 0 \ea
\] which means that the reduced density matrix between any two
spins has the particularly simple form \cite{EntRings}
\[ \rho_{12} = \left( \ba{cccc} v & & & \\ & w & z & \\ & \bar{z} & x & \\ & & & y \ea \right) \]
in the basis $ \left\{ \dket{00}, \dket{01}, \dket{10}, \dket{11}
\right\} $.  The concurrence for such a density matrix is given by
\cite{EntRings}
\begin{equation}
C = 2\max\{|z| - \sqrt{vy},0\}. \label{concexpression}
\end{equation}
In \cite{starspin} we used this formula to obtain an analytic
formula for the concurrence between any two outer spins in the
star model
\begin{equation}
\ba{l}
C = 2\max\{1/2N,0\} = 1/N \text{ for $N$ odd} \\
C = 2\max\{1/2N - 1/(2N^2-2N),0\} = 1/N- 1/(N^2-N) \text{ for $N$ even}
\ea
\label{eqn:srconcs}
\end{equation}

To compare with the above, we numerically evaluate the concurrence
for the ring for a range of values of $N$. Wang \cite{XXRings} has
ascertained that for \emph{even} $N$ the concurrence is
independent of the sign of the coupling constant ${\cal J}$ which
multiplies the Hamiltonians (the Hamiltonians being ${\cal
J}H_{\text{star}}$ and ${\cal J}H_{\text{ring}}$. We note that
this is also true for both even and odd $N$ for the star model
(the formulae in \ceqn{eqn:srconcs} do not depend on ${\cal J})$.

The one major advantage that the star has, just by shear virtue of
its geometry, is that the entanglement between any two outer spins
is the same as there is perfect permutation symmetry in the model.
This fact tells us that the star model will most probably be
superior in comparison to the ring model for sharing entanglement
between spins that are not necessarily physically adjacent ({\em
i.e.}, nearest neighbors in the sense of a ring).


In section \ref{sec:compsr} we directly compare the star and ring models for sharing entanglement between neighbouring spins.  Then in section \ref{sec:invcombo} we investigate how successful a model in which both interactions occur is at sharing entanglement.
An insight into the reason for the behaviour of the combination is given by the energy level crossings which is highlighted in section \ref{sec:elevelcrossings}.  An full explanation for the case of $N=4$ is presented in section \ref{sec:expln4} and section \ref{sec:generalbonds} makes some generalisations for any $N$.  Finally we present our conclusions in section \ref{sec:srcomboconc}.

\section{Comparing the star and ring spin models \label{sec:compsr}}


In this section we compare how well the ground state of the star and ring models share entanglement between neighbouring spins.  For both models we find the density matrix numerically for the ground state for $2 \leq N \leq 6$ and select two spins by tracing out remaining spins from the density matrix.  The concurrence between the two spins is then used as a meaure of the entanglement between them.

\begin{figure}
\begin{tabular}{cc}
\includegraphics[width=2.85in, clip]{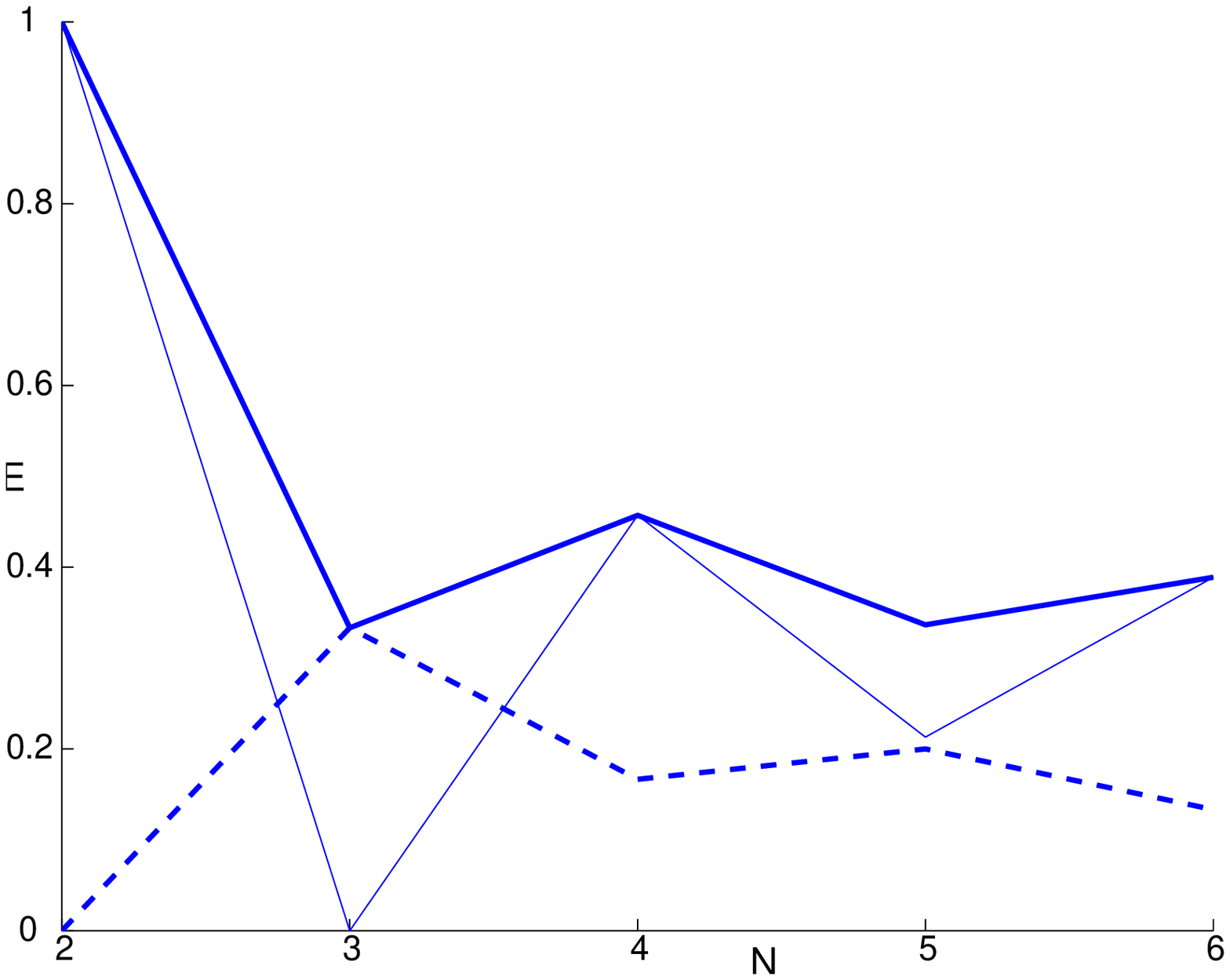} & \includegraphics[width=2.85in, clip]{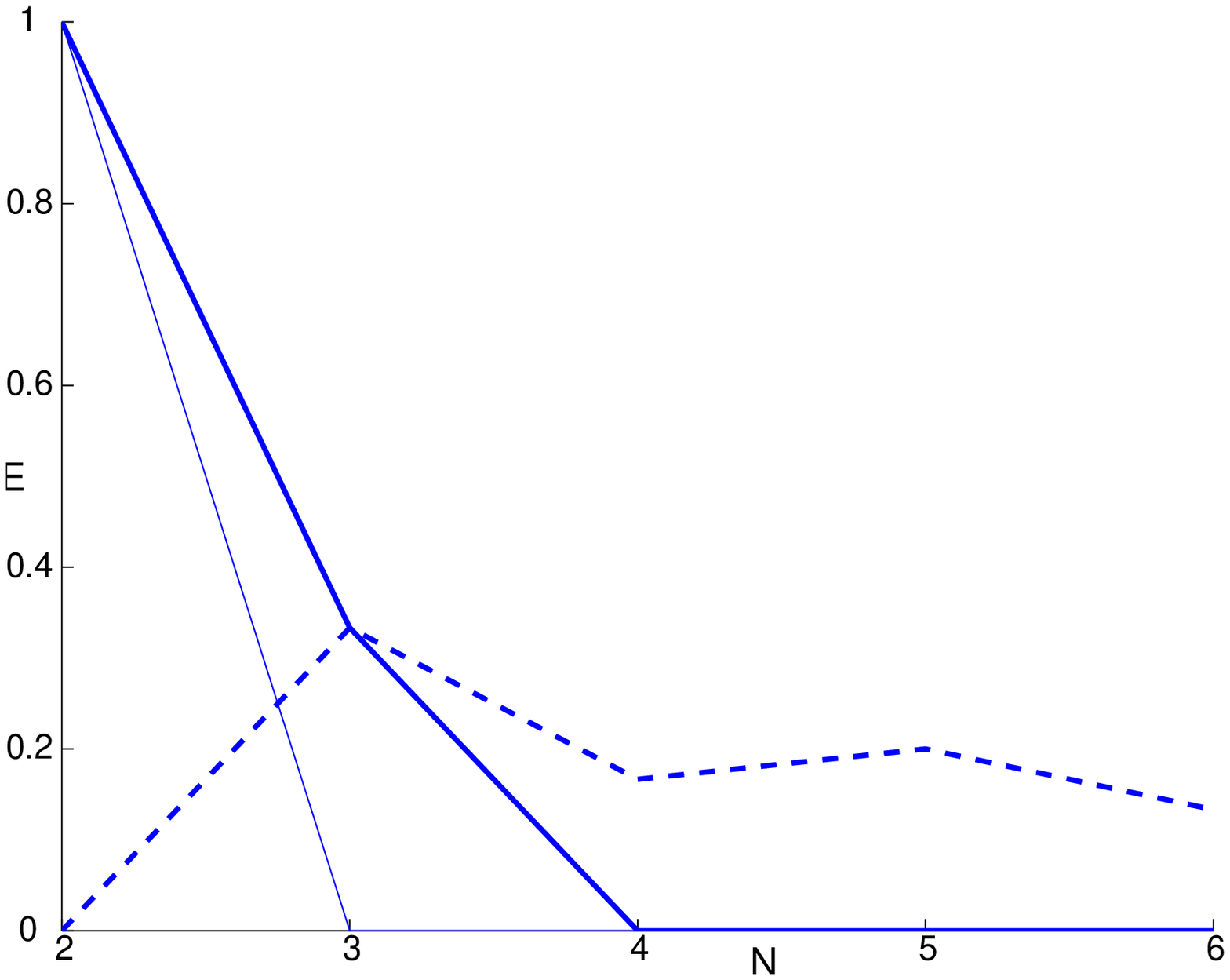} \\
(a) & (b) \\
\end{tabular}
\caption{(a) This figure plots the nearest neighbour concurrence.
(b) This figure plots the next to nearest neighbour concurrnce.
The solid line is the ring and the dashed line is the star (for
the star, nearest and next to nearest neighbors would be defined
in an ad-hoc basis, in terms of physical adjacency, rather than in
terms of interactions). The thin lines depict the
antiferromagnetic interaction and the thick lines depict the
ferromagentic interaction. Note that for the star, the thick and
thin lines are coincident} \label{fig:srnnntn}
\end{figure}
\cfig{fig:srnnntn}(a) plots the nearest neighbour entanglement of
two outer spins.  We observe that the entanglement of the star is
not affected by whether the interaction is ferromagnetic or
antiferromagnetic.  Note that for $N=2$ and $N=3$ all the spins
are nearest neighbours.  Also note that for even $N$, the ring is
not affected by the sign of $J$, which agrees with Wang in
\cite{XXRings}.

\cfig{fig:srnnntn}(b) plots the next to nearest neighbour
entanglement. The most noticeable feature of this graph is that
the ring entanglement drops to zero and completely vanishes. In
fact, for $N=2$ and $N=3$ the points on the graph really represent
nearest-neighbour entanglement.  Therefore we can venture to
conjecture that there is \emph{no} next to nearest entanglement in
the ring model. On the other hand, the star displays exactly the
same behavior for any pair of spins, as would be expected from the
symmetry of the model.

In summary then, for nearest neighbour interactions the ring appears (in the limited range of $N$ considered) to have slightly higher entanglement than the star, but not very dissimilar for $N$ odd, and furthermore the star model shares entanglement between states for next to nearest neighbours whereas the ring shares none at all.  Finally we also note that the ferromagentic interaction is the best for sharing entanglement in the ring (and equally good in the star as the antiferromagnetic interaction).

\section{Combining the star and ring spin models \label{sec:invcombo}}

In the previous section, we compared the star and ring models for
establishing entanglement between spins. We observed that the ring
was superior at establishing nearest-neighbour entanglement while
the star was better for establishing next-to-nearest neighbour
entanglement. It is then then natural to wonder about the nature
of entanglement between spins in a model in which the spins could
interact both with the nearest neighbour spins (like in the ring)
and with a central spin (like in a star). This is what we
investigate in rest of the paper. In this section, we will find
out whether the combination of star and ring interactions can
produce an entanglement (between nearest neighbor pairs or
non-nearest neighbor pairs) which is higher than that of the star
alone or the ring alone. The Hamiltonian for such a system could
be written
\begin{equation}
H = {\cal J} \left[ c H_{\text{star}} + (1-c) H_{\text{ring}} \right]
\label{eqn:combosr}
\end{equation}
where $c$ is a parameter $0 \leq c \leq 1$ which interpolates between the two extremes of the ring ($c=0$) and the star ($c=1$).

\begin{figure}[t]
\begin{tabular}{cc}
\includegraphics[width=2.85in, clip]{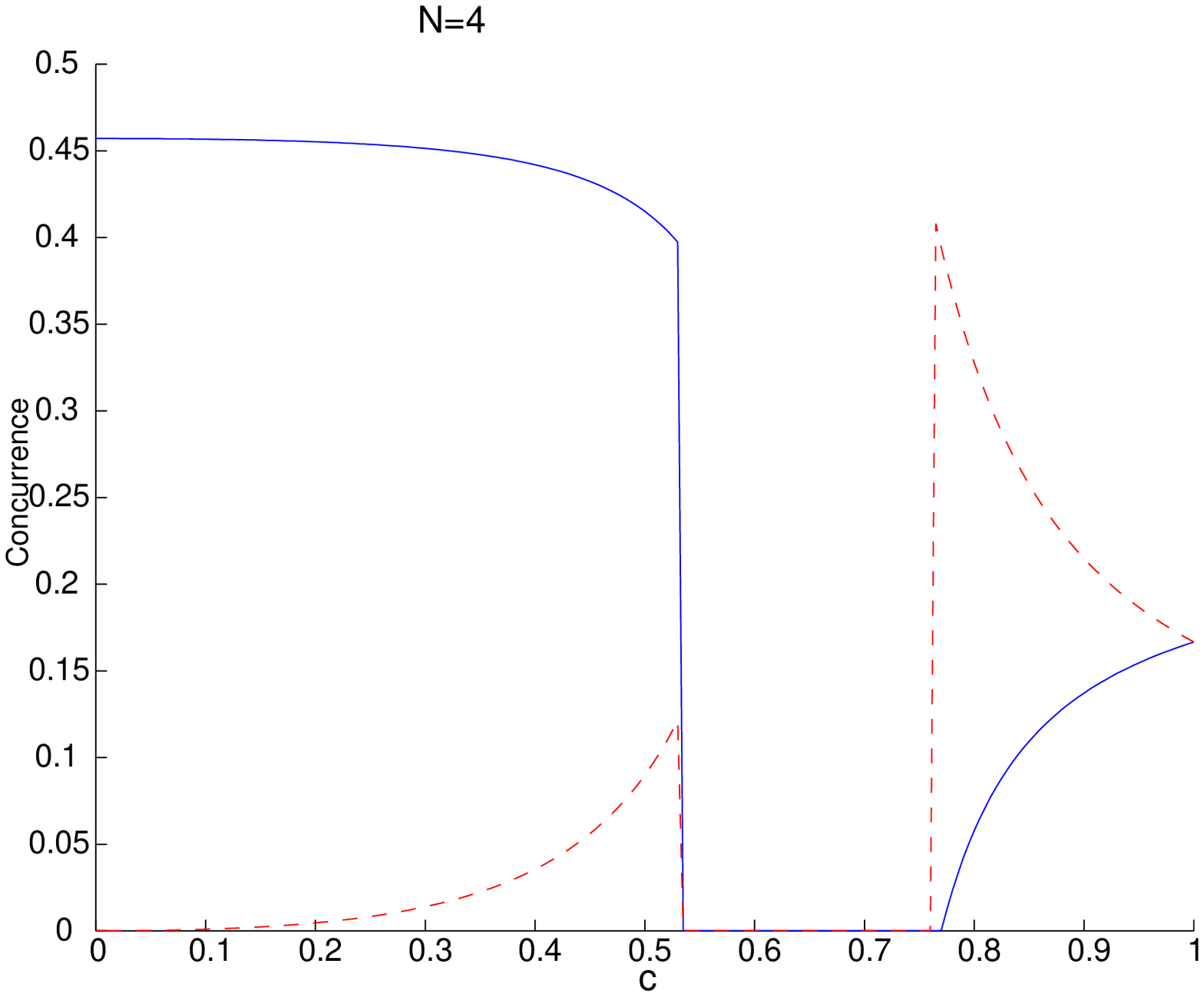} &
\includegraphics[width=2.85in, clip]{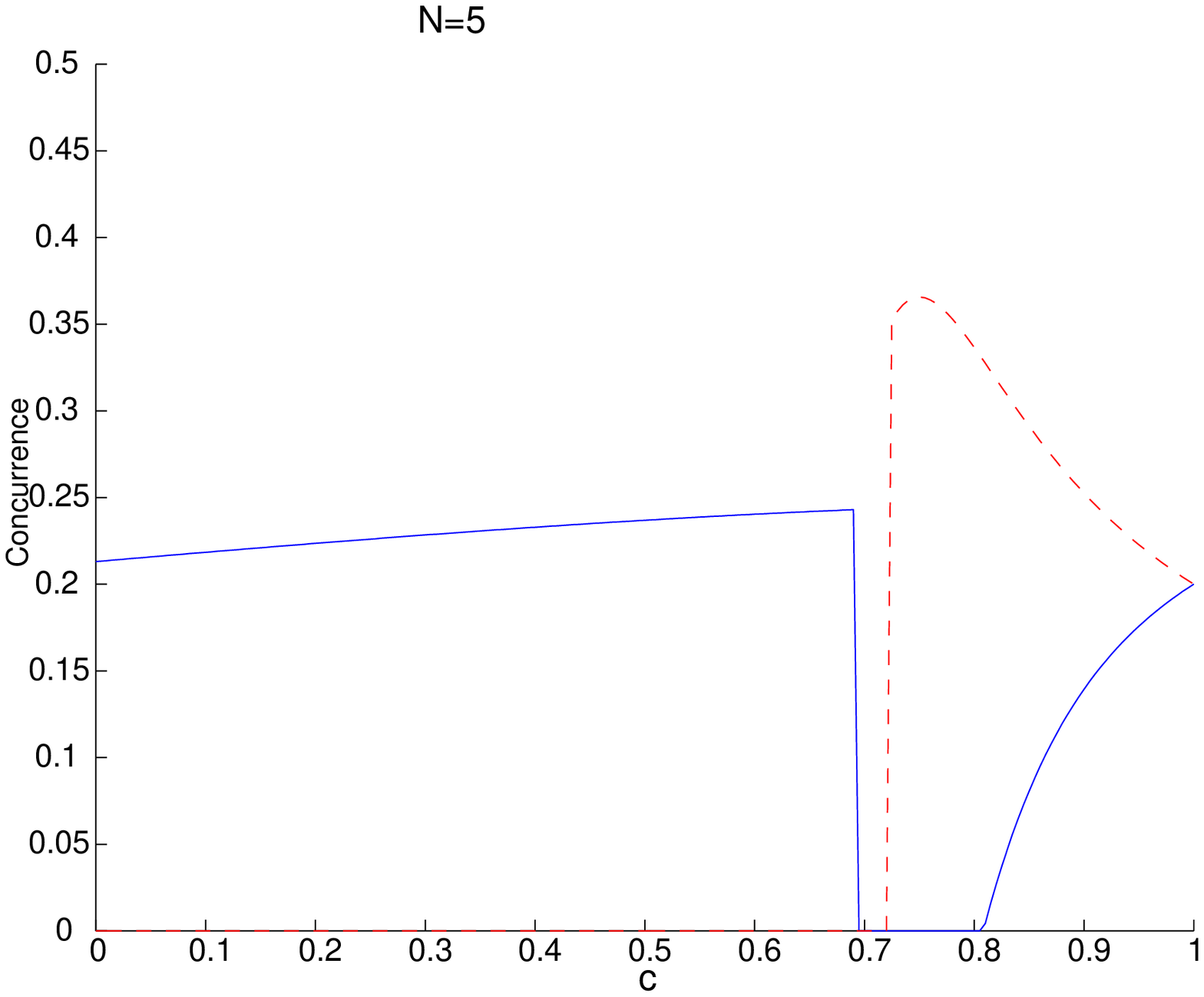}
\\
\includegraphics[width=2.85in, clip]{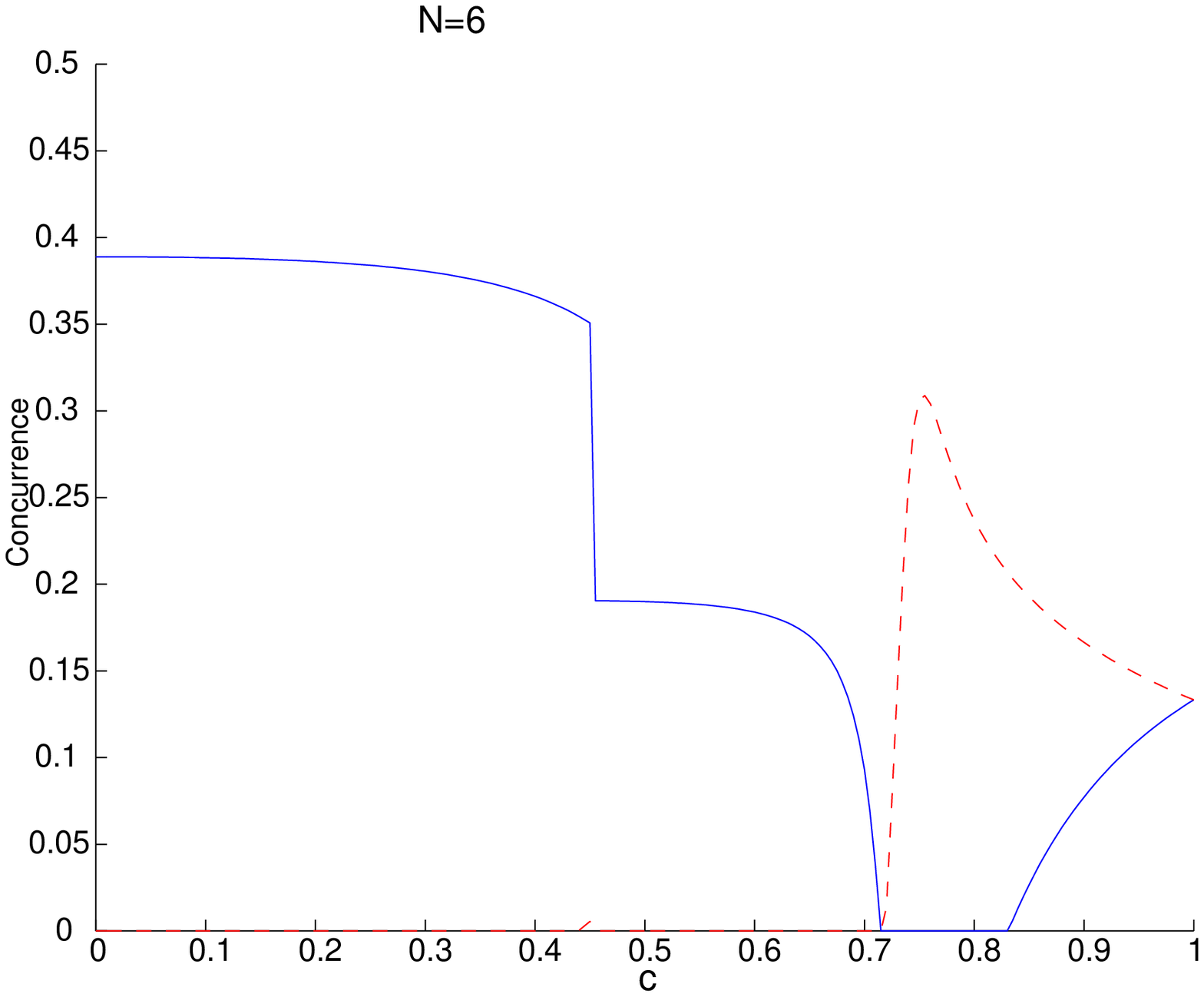} &
\includegraphics[width=2.85in, clip]{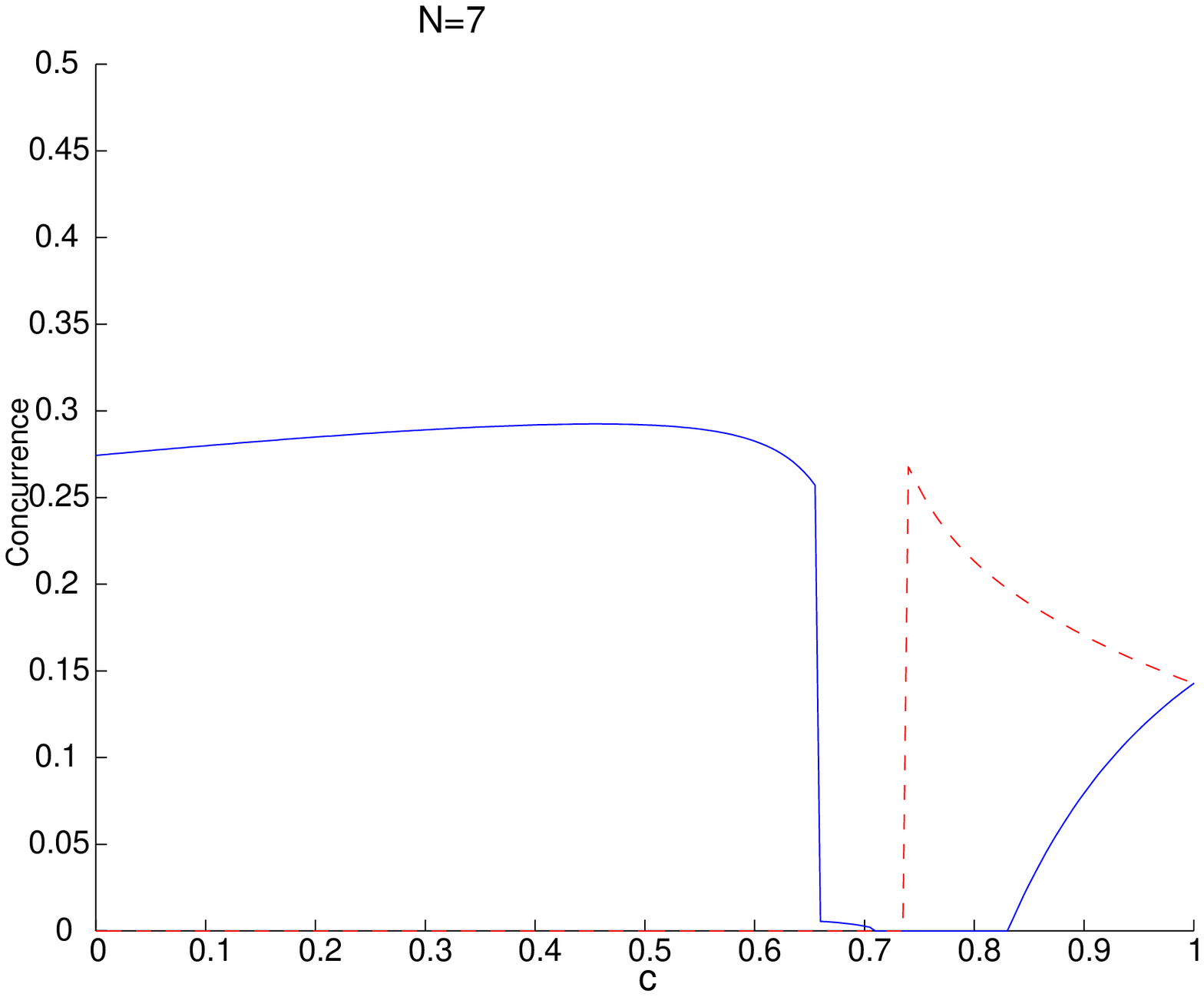}
\\
\end{tabular}
\caption{The ground state entanglement for different $n$ as $c$ varies from 0 (ring)
to 1 (star).  The solid line and the dashed line depict nearest neighbour and next-to-nearest neighbour entanglement respectively.} \label{fig:gent}
\end{figure}

Our interest is in how the concurrence varies with $c$ and whether
it is maximal for a non-extremal value of $c$. While there are
well known analytical methods for solving for the ground state for
both the extremal values of $c$ (one can solve the ground state of
$H_{\text{ring}}$ by using the Bethe Ansatz \cite{Bethe} and we
have already presented the ground state of $H_{\text{star}}$ in
Ref.\cite{starspin}), there does not seem to be any easy to apply
analytic technique for obtaining the ground state for arbitrary
$c$. So we have obtained the ground state for the Hamiltonian in
\ceqn{eqn:combosr} numerically and calculated the concurrence
between two of the spins lying on the ring. \cfig{fig:gent} plots
the concurrence against $c$ for $N=4,5,6$ and $7$ for both
nearest-neighbor (solid line) and next to nearest neighbor (dashed
line) concurrence. These values of $N$ were chosen for
computational convenience. There are a number of interesting
observations that can be made at once from these plots:
\begin{itemize}
\item For odd $N$, there appears to be an initial rise in
nearest-neighbor entanglement as $c$ increases from $0$. \item For
any $N$, the maximum next-to-nearest neighbor entanglement occurs
at a value of $c$ less than $1$, i.e.\ before the network becomes
entirely star-like. \item There are a number of sharp changes in
the entanglement. \item In general there seems to be a point
around $c=0.7$ where all entanglement drops to zero.
\end{itemize}
This suggests that a combination of the two models can maximize
either the nearest neighbor or the next-to-nearest neighbor
entanglement.  We use `either \ldots or' because the two maxima do
not occur at the same value of $c$. Thus, though a ring
interaction ($c=0$) does not, by itself, favor next to nearest
neighbor entanglement (its value being zero), it can be mixed with
the star to actually {\em increase} the next to nearest neighbor
entanglement of the star system. This is a counterintuitive
feature. Moreover, for odd $N$, a proportion of the star
interaction seems to increase the nearest neighbor entanglement,
though the star is expected to remove the special status of
nearest neighbors of a ring system. Furthermore we observe that
there are sharp jumps in entanglement, which include drops to zero
at an or for a range of intermediate values of $c$. These features
are consequently quite surprising and we will devote majority of
the rest of the paper to seeking their explanation.

\begin{figure}
\begin{tabular}{cc}
\includegraphics[width=2.85in, clip]{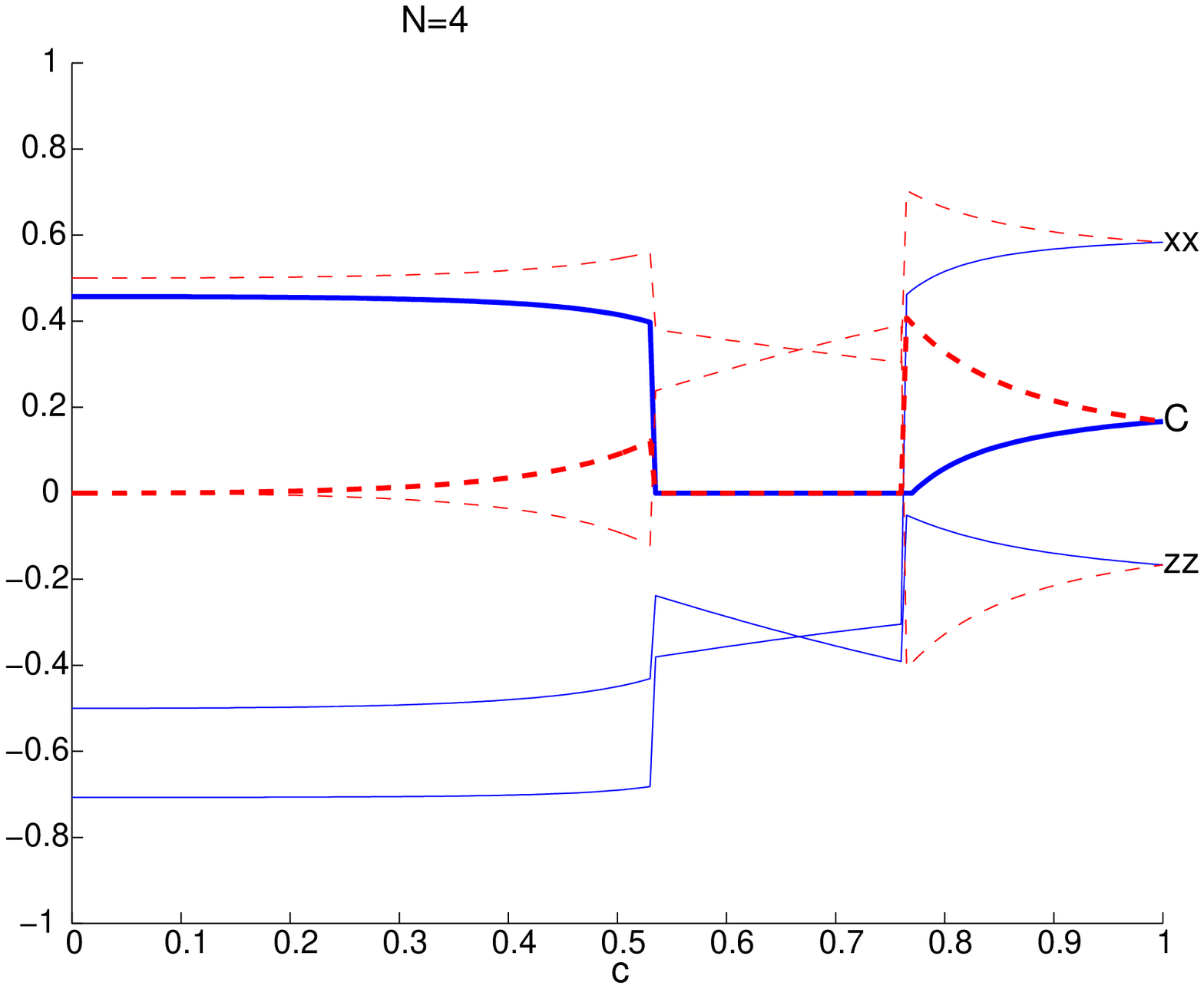} &
\includegraphics[width=2.85in, clip]{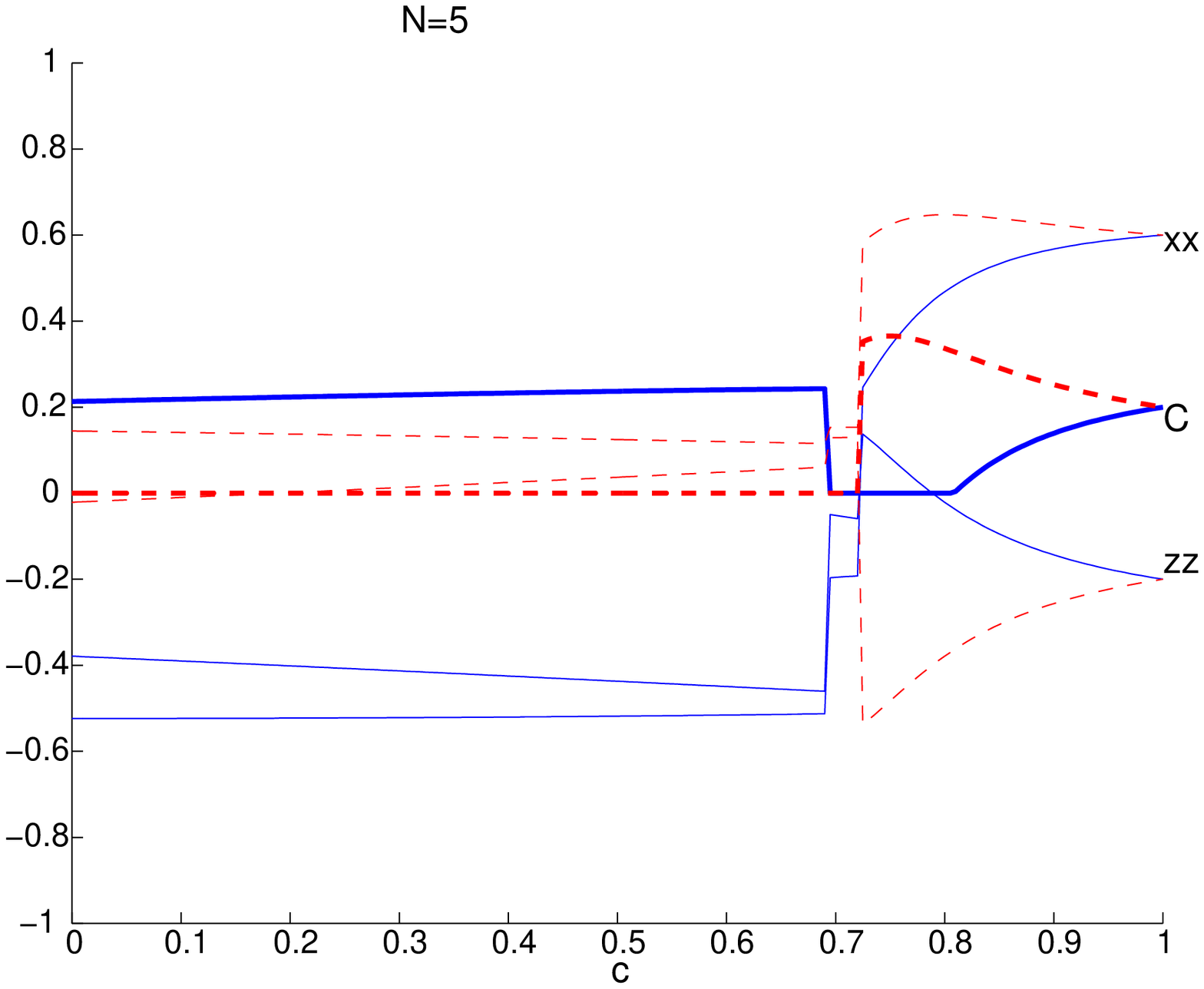}
\\
\includegraphics[width=2.85in, clip]{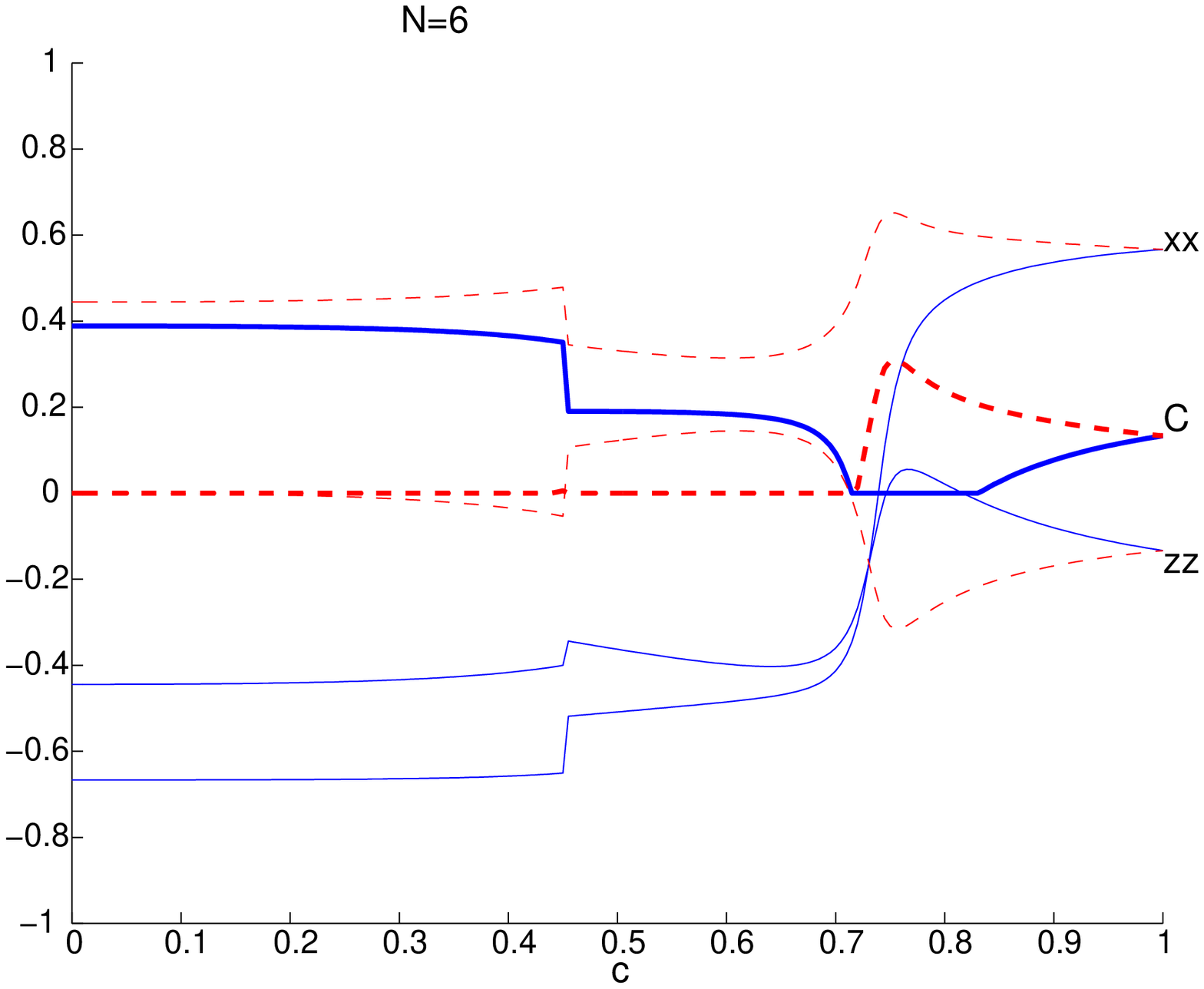} &
\includegraphics[width=2.85in, clip]{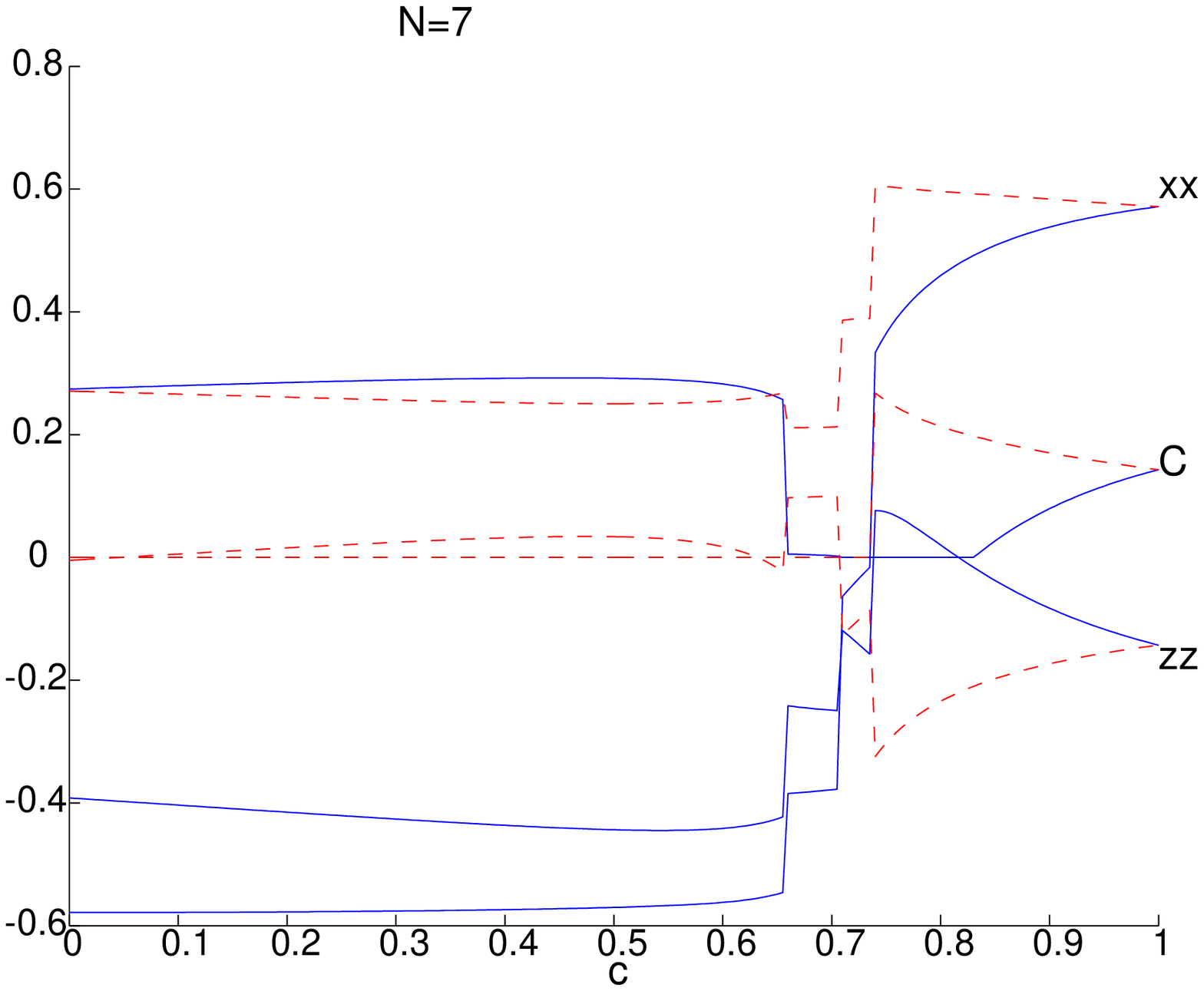}
\\
\end{tabular}
\caption{$XX$ and $ZZ$ correlations in the ground state for different $n$ as $c$ varies from 0 (ring)
to 1 (star).  Also shown in the concurrence (the thicker lines).  The solid line and the dashed line depict nearest neighbour and next-to-nearest neighbour entanglement respectively.} \label{fig:gcorrels}
\end{figure}

 Having looked at the concurrence, we now also take a brief look at the localizable entanglement in the
combination of the star and the ring models.  The localisable
entanglement has been introduced by Verstraete et al.\
\cite{Localizable} and studied further by others \cite{jin}, and
is the average amount of entanglement that can be established
between two spins by performing local measurements on the other
individual spins. The authors of Ref.\cite{Localizable} showed
that all classical correlation functions provide lower bounds to
this localizable entanglement. In \cfig{fig:gcorrels} we have
plotted the $XX$ and $ZZ$ correlations (which provide lower bounds
on the localizable entanglement ) together with the concurrence
that was plotted in \cfig{fig:gent}. There are two main points of
interest. Firstly the correlations (and hence the lower bound on
the localizable entanglement) are non-zero even in areas where the
concurrence is zero, in particular around $c\approx 0.7$. Thus
these areas are not uninteresting in terms of entanglement. In
fact, in these regimes of $c$, one can search for interesting
multiparticle entangled states (two particle entanglement being
zero does not imply that the system of spins in not in a
multiparticle entangled state). Secondly, we note that the
localizable entanglement is quite high in general for most values
of $c$, both for nearest neighbors and next to nearest neighbors.
For example, for $N=4$ it ranges from $0.7$ for the ring to $0.6$
for the star, dropping to a low of $0.4$ for intermediate values
of $c$.  Therefore the localizable entanglement in this model is
significant in magnitude.

\begin{figure}[t]
\begin{tabular}{cc}
\includegraphics[width=2.85in, clip]{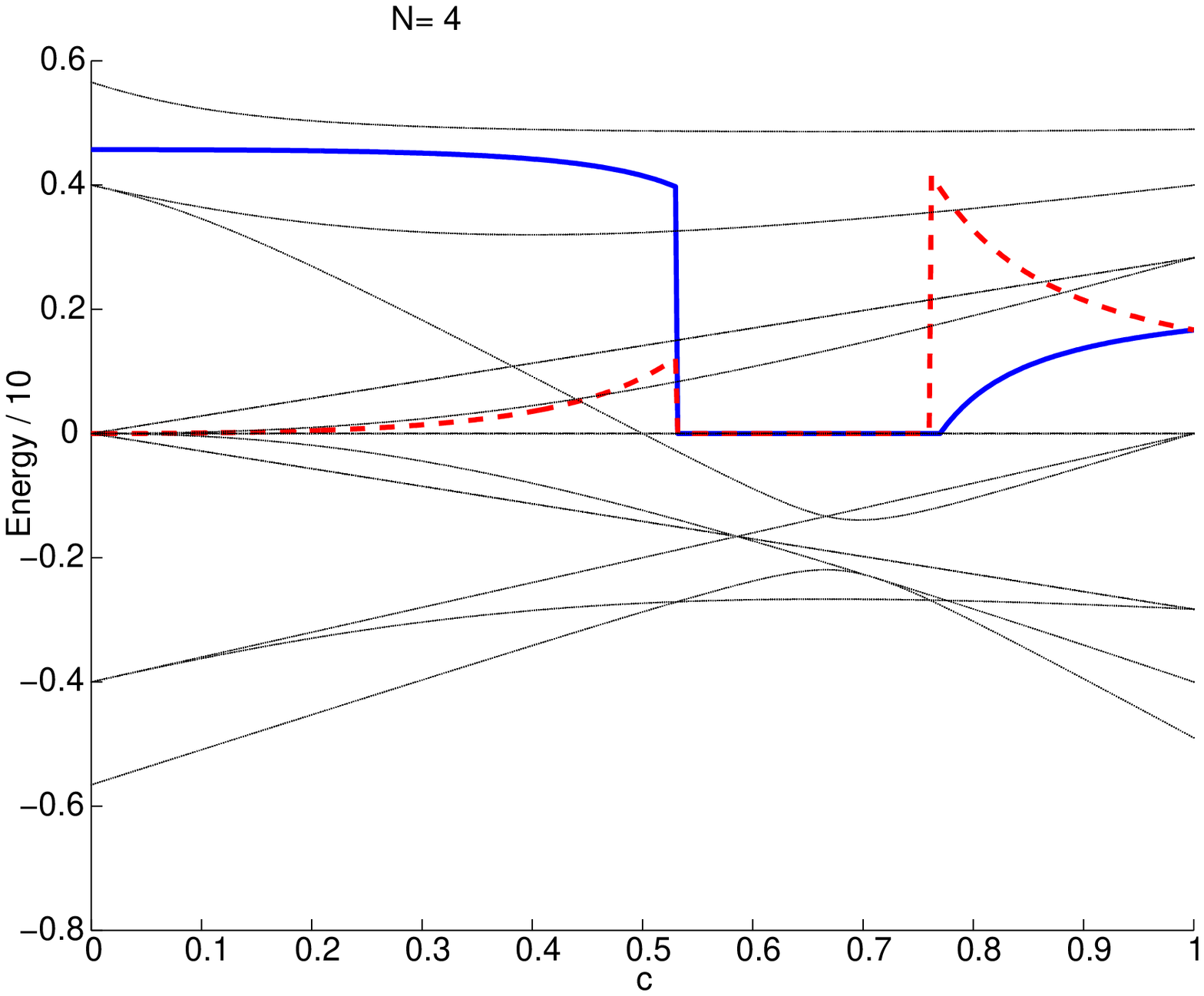} &
\includegraphics[width=2.85in, clip]{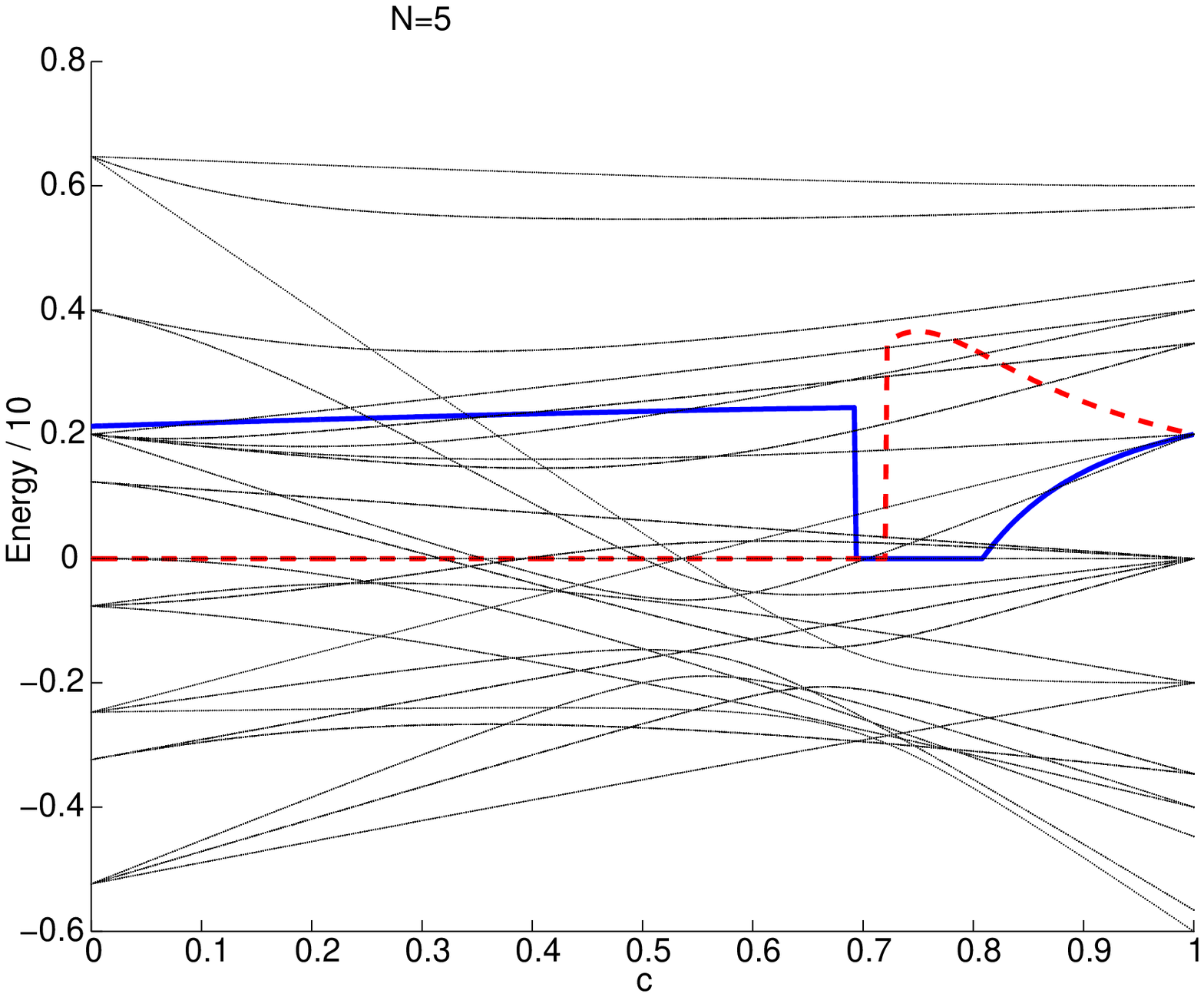}
\\
\includegraphics[width=2.85in, clip]{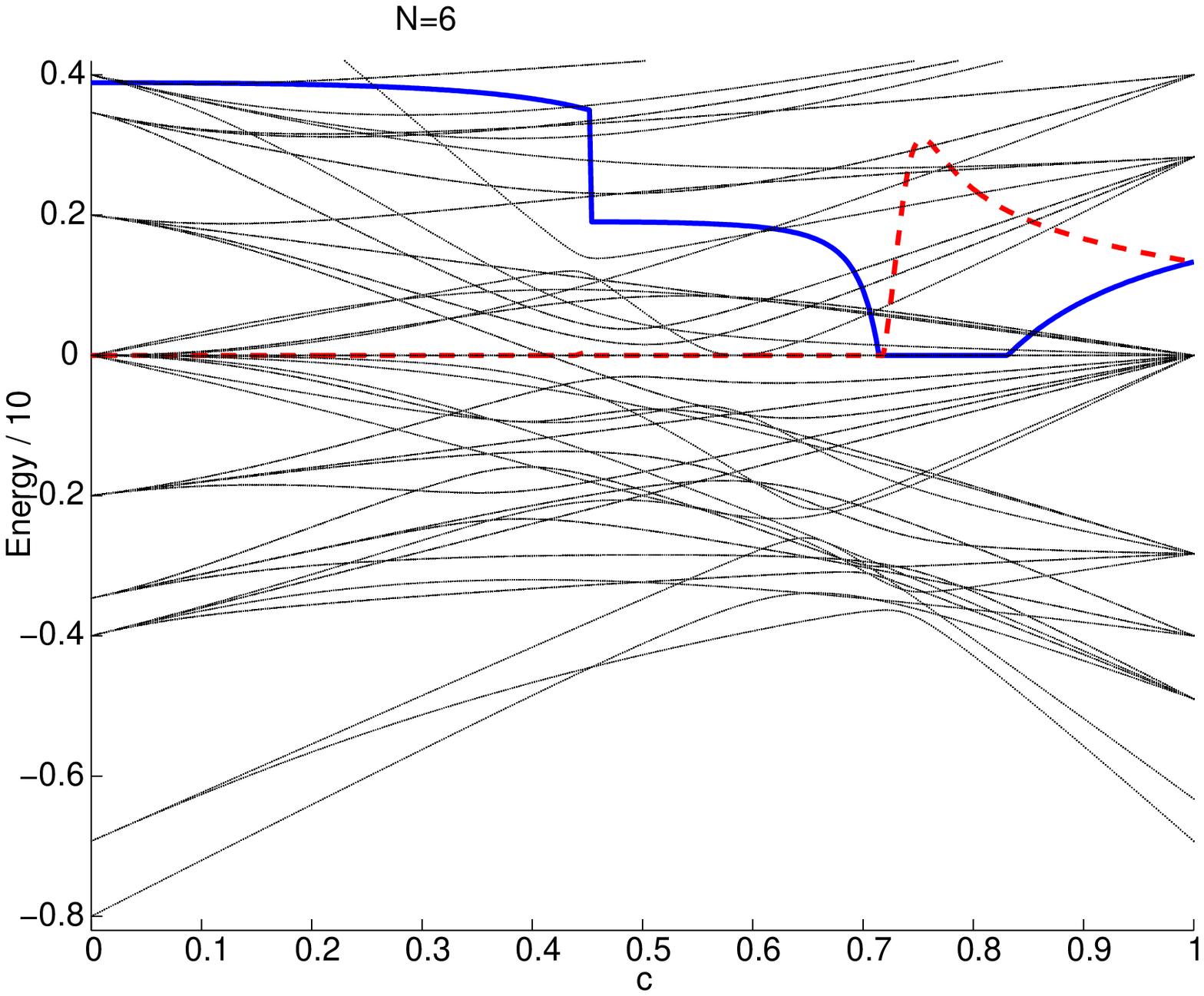} &
\\
\end{tabular}
\caption{The energy levels for different $n$ as $c$ varies from 0
(ring) to 1 (star).} \label{fig:energies}
\end{figure}
%

\section{Energy level crossings and the sharp changes in entanglement \label{sec:elevelcrossings}}

Some light can be shed on the reason behind one of the results
described in the previous section (namely the sharp changes in
concurrence as $c$ is varied) by studying the energy levels of the
system and how they vary with $c$. Energy level diagrams are
plotted in \cfig{fig:energies}. Superposed on the diagrams are the
nearest and next to nearest neighbor concurrences.  The figure
clearly shows that the jumps in the concurrence are due to
crossings of the lowest energy levels. Thus the sharp transitions
in entanglement are in one to one correspondence with a sudden
qualitative change in the ground state of the system which happens
due to level crossings. The sudden change in the ground state
energy level changes the two spin reduced density matrices and
thereby the concurrence. Such changes are also the cause of
quantum phase transitions, which occur in infinite systems. Here
we note the similarity, though our system is finite. In other
words, the cause of quantum phase transitions (competing
Hamiltonian terms causing energy level crossings or infinitesimal
avoided level crossings), when applied to our finite system, also
causes sharp transitions in entanglement.

\section{The variation of eigenstates from star to ring \label{sec:changingeig}}

\begin{figure}
\begin{tabular}{cc}
\includegraphics[width=2.85in, clip]{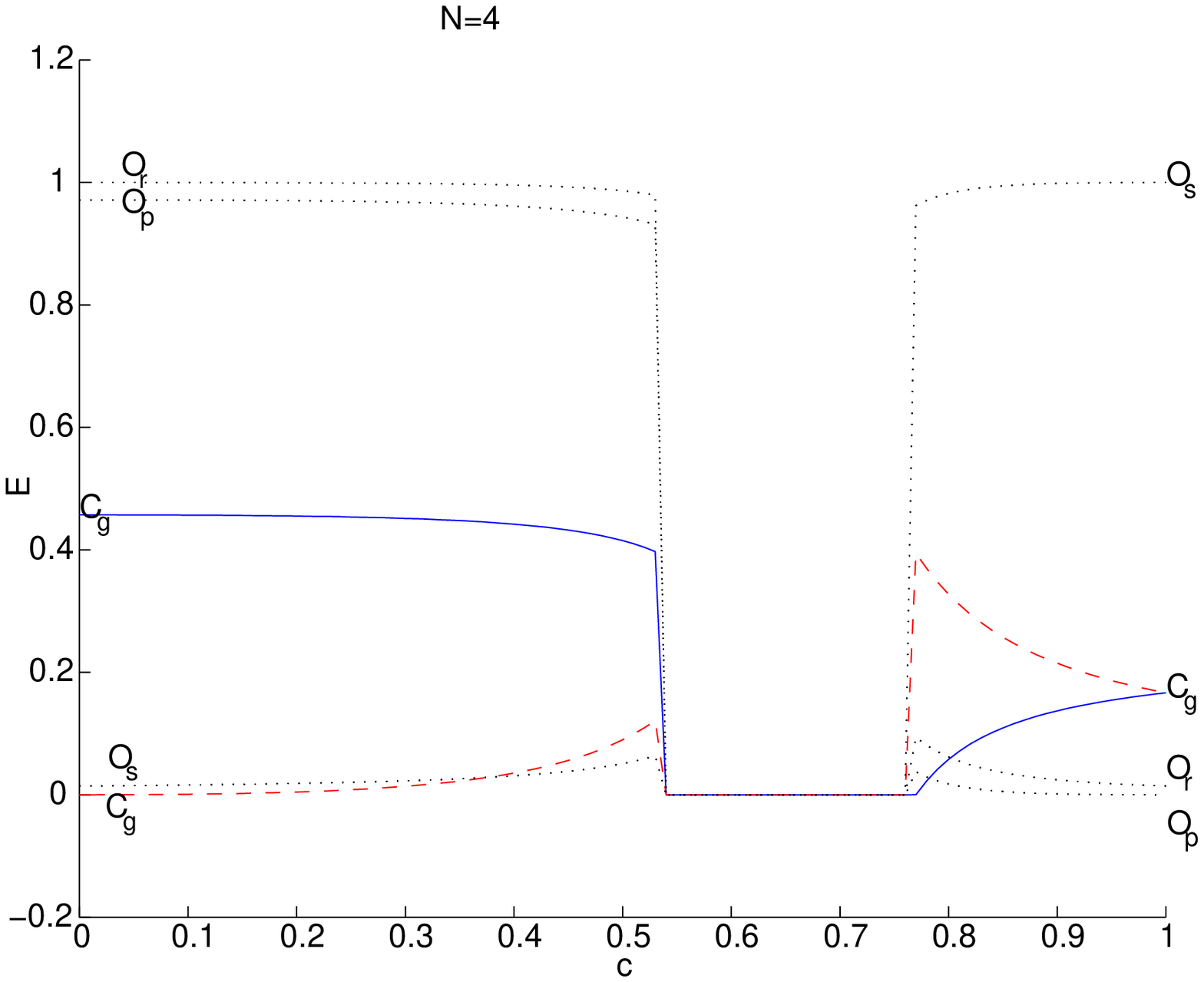} &
\includegraphics[width=2.85in, clip]{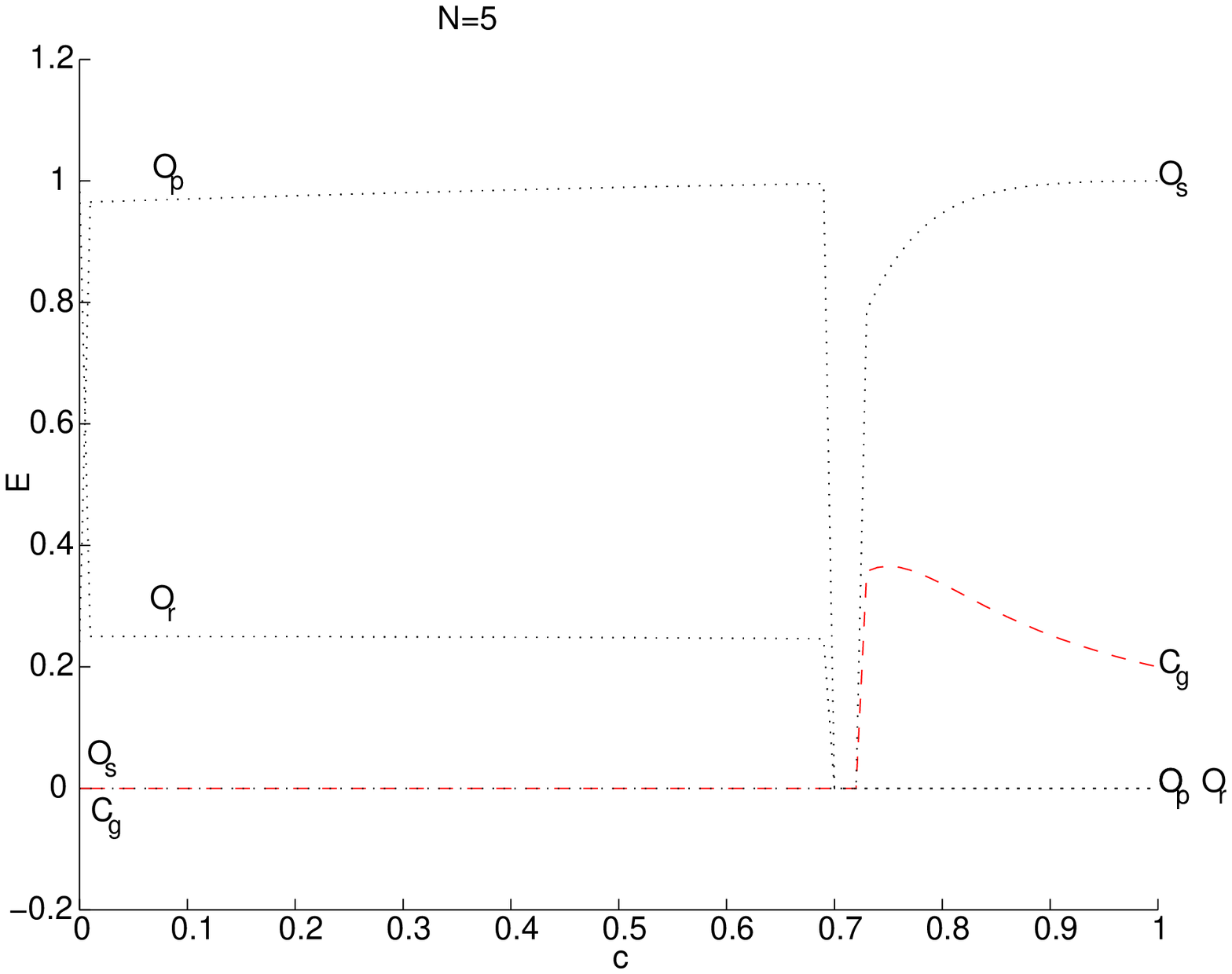}
\\
\includegraphics[width=2.85in, clip]{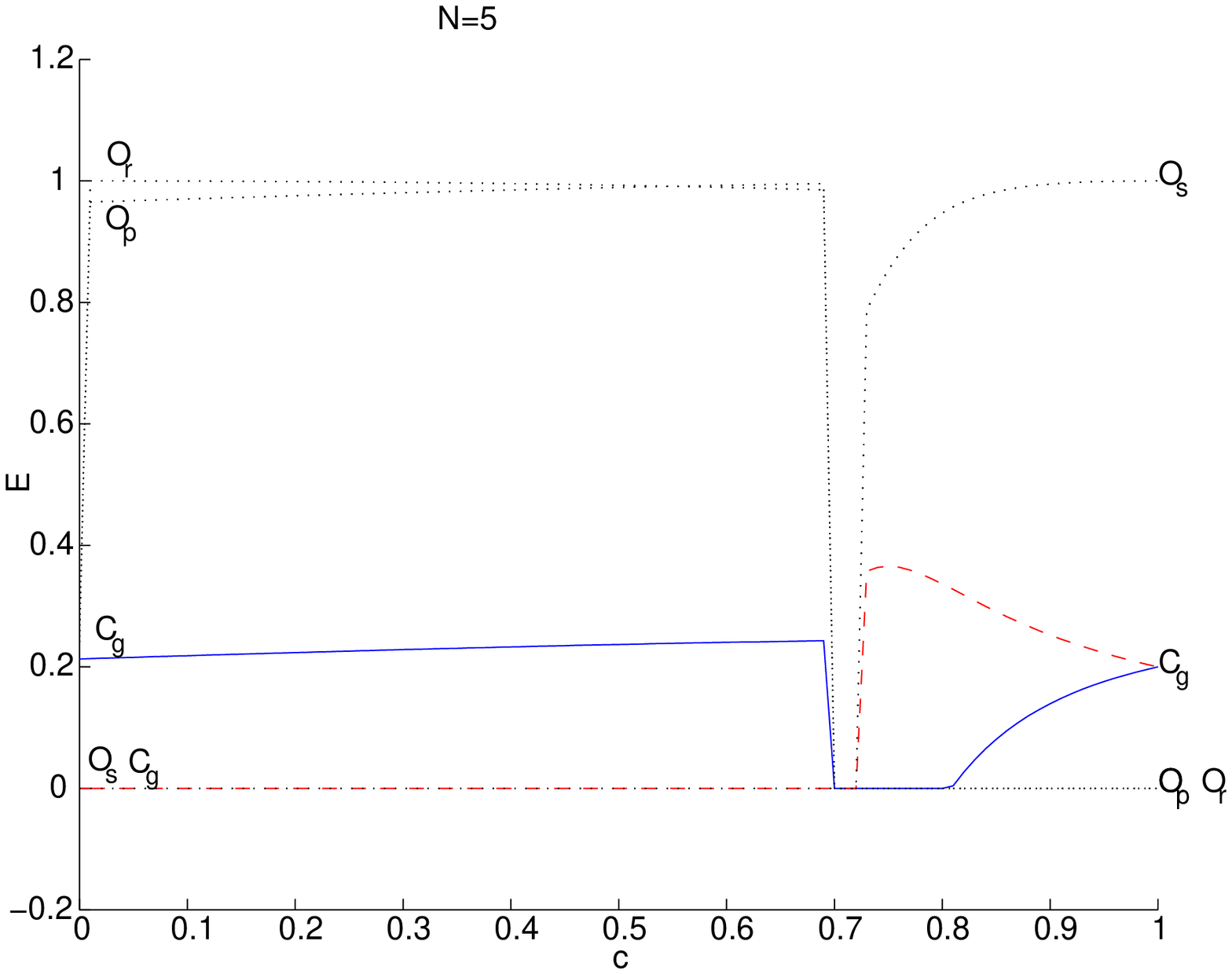} &
\includegraphics[width=2.85in, clip]{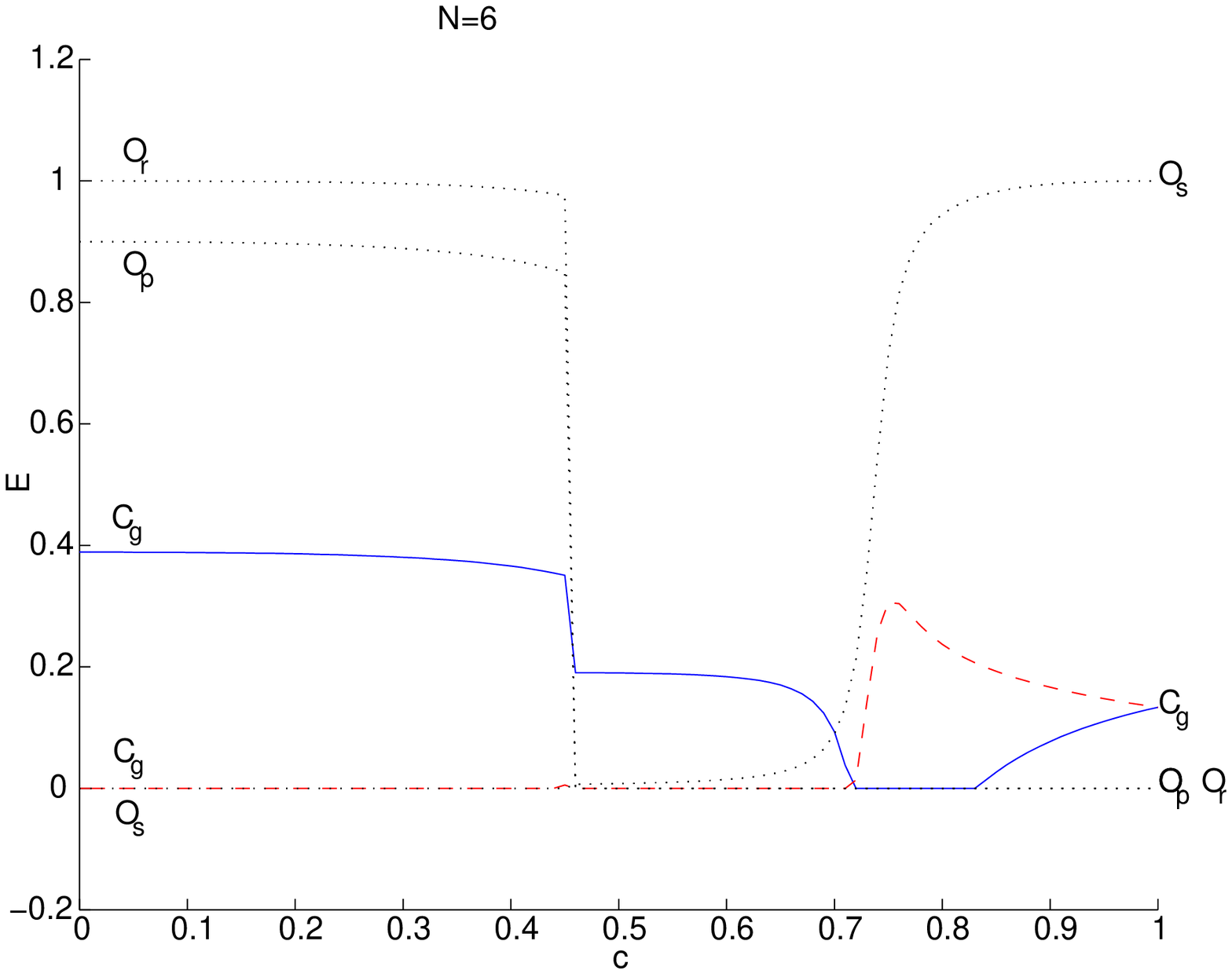}
\\
\end{tabular}
\caption{This figure plots the fidelity of the ground state at a
certain value of $c$ with three states.  $O_p$ is the fidelity
with a speculated state composed of singlets.  $O_r$ is the
fidelity with the ground state of the pure ring.  $O_s$ is the
fidelity with the ground state of the pure star.  Also plotted
here are the concurrences for reference, marked by $C_g$.}
\label{fig:ovs}
\end{figure}

An initial impression of how the ground states of the model change as $c$ varies can be obtained by studying how similar they are to the pure ring ($c=0$) and pure star ($c=1$) ground states.  The ground states can be compared by calculating the fidelity $F$ \cite{Fidelity} of the two density matrices for the two cases.  The fidelity is a measure of `how close' two states $\rho_1$ and $\rho_2$ are.
It ranges in value from $0$ to $1$ and is equal to one if and only if $\rho_1$ and $\rho_2$ are equal.  It is defined by
\begin{equation}
F(\rho_1, \rho_2) = \left[ \text{tr} \sqrt{ \sqrt{\rho_1} \rho_2 \sqrt{\rho_1} } \right]^2
\end{equation}
In \cfig{fig:ovs} the fidelity of the ground state as $c$ varies with the pure star and pure ring ground states has been plotted for $N=4,5$ and $6$.  The fidelity with the ring ground state has been labelled by $O_r$ and with the star ground state labelled by $O_s$.  If we denote the ground state by $\rho_g$ then
\begin{eqnarray*}
O_r & = & F \left[ \rho_g(c), \rho_g(c=0) \right] \\
O_s & = & F \left[ \rho_g(c), \rho_g(c=1) \right]
\end{eqnarray*}
What is illustrated by \cfig{fig:ovs} is that the eigenstates
remain `ring-like' and `star-like' for much of the range of $c$
either side of $c \approx 0.7$.  This is shown by the fact that
$O_r$ is near to one for much of $c<0.7$ and $O_s$ is near to one
for much of $c>0.7$ (the other overlap $O_p$ appearing in
\cfig{fig:ovs} is introduced and discussed in section
\ref{sec:generalbonds}).

Note that for the case of $N=5$ the overlap with the ring $O_r$
may seem very low, as it is approximately 0.2 rather than the 1
that might be expected.  This is because the ground state of the
pure ring is 8-fold degenerate.  This degeneracy is lifted for $c
\neq 0$, and therefore its overlap with states for which $c \neq
0$ will be smaller.  In the bottom left plot in \cfig{fig:ovs} is
the same graph, but this time $O_r$ is the fidelity not with the
pure ring, but with the state for which $c=0.01$.  This was done
to lift the degeneracy in the `pure ring' state used to calculate
the fidelity.

The $N=5$ case then confirms the same pattern as was observed for $N=4$ and $N=6$.  This is that the ring and star ground states are quite stable, because the ground state stays close to that of the ring (near $c=0$) or the star (near $c=1$) for a considerable range of values of $c$.

In summary then, these initial numerical investigations for $N=4,5,6$ and $7$ indicate that there is interesting structure present in a combination of the star and ring models and that it deserves further study.  The next section describes a detailed study of the cases for $N=4$.

\section{$N=4$ in detail \label{sec:expln4}}

In this section, we give a mathematical explanation for the
behavior of the entanglement in the ground state for the case
$N=4$ by presenting the exact eigenstates for this case (a more
physical explanation will be presented in the discussions
section). The energy level diagram in \cfig{fig:energies} for
$N=4$ indicates that there are two different energy levels which,
depending on the value of $c$, are the ground state.  The energy
level which is the ground state for most of the time, except for
$c$ near 0.7, is the ground state for both the pure ring and pure
star.  We call this energy level I, and the other energy level,
which is the ground state only for a short while near $c=0.7$, we
call energy level II.  Our aim in this section is to analyze how
the concurrence varies for these two energy levels and to
determine why one energy level takes over from the other as the
ground state for a certain range of $c$.  We will begin by setting
out for reference a number of relevant states and their actions
under the Hamiltonians being studied. After reviewing the two
energy levels of interest for the extremes of the star and ring
models, we will describe the behavior of the energy levels for
$0<c<1$.

Firstly, we define some convenient states which will be frequently used:

\[ \ba{lcl}
\dket{A} & = & \rt \left( \dket{0101} + \dket{1010} \right) \\
\dket{B} & = & \frac{1}{2} \left( \dket{0011} + \dket{0110} + \dket{1100} + \dket{1001} \right) \\
\dket{C_1} & = & \frac{1}{2} \left( \dket{0001} + \dket{0010} + \dket{0100} + \dket{1000} \right) \\
\dket{C_3} & = & \frac{1}{2} \left( \dket{0111} + \dket{1011} + \dket{1101} + \dket{1110} \right) \\
\dket{C_1'} & = & \frac{1}{2} \left( \dket{0001} - \dket{0010} + \dket{0100} - \dket{1000} \right) \\
\dket{C_3'} & = & \frac{1}{2} \left( \dket{0111} - \dket{1011} + \dket{1101} - \dket{1110} \right) \\
\dket{D} & = & \rt \left( \dket{0101} - \dket{1010} \right) \\
\ea \]

Note that these states are rotationally invariant.  This is because the ring is rotationally invariant.  In fact so is the star, although the star also has the stronger permutation symmetry.  As the eigenstates of the star are made up from angular momentum eigenstates
\cite{starspin},
the following relations will also be useful

\begin{eqnarray*}
\dket{j=2,m=0} & = & \frac{1}{\sqrt{6}} \left( \sqrt{2} \dket{A} + 2\dket{B} \right) \\
\dket{j=1,m=1} & = & \dket{C_3'} \\
\dket{j=1,m=0} & = & \dket{D} \\
\dket{j=1,m=-1} & = & \dket{C_1'} \\
\end{eqnarray*}

Table \ref{tbl:sraction} shows how these states (plus the central spin) are affected by the star and ring Hamiltonians individually.

\begin{table}
\begin{center}
\begin{tabular}{l@{\extracolsep{1.5em}}ll}
\hline
& $H_{\text{star}}$ & $H_{\text{ring}}$ \\
\hline
$\dket{0}\dket{A}$ & $2\sqrt{2} \dket{1}\dket{C_1}$ & $4\sqrt{2} \dket{0}\dket{B}$ \\
$\dket{1}\dket{A}$ & $2\sqrt{2} \dket{0}\dket{C_3}$ & $4\sqrt{2} \dket{1}\dket{B}$ \\
$\dket{0}\dket{B}$ & $4\dket{1}\dket{C_1}$ & $4\sqrt{2} \dket{0}\dket{A}$ \\
$\dket{1}\dket{B}$ & $4\dket{0}\dket{C_3}$ & $4\sqrt{2} \dket{1}\dket{A}$ \\
$\dket{0}\dket{C_1}$ & $4\dket{1}\dket{0000}$ & $4\dket{0}\dket{C_1}$ \\
$\dket{1}\dket{C_1}$ & $2\sqrt{6} \dket{0}\dket{j=2,m=0}$ & $4\dket{1}\dket{C_1}$ \\
$\dket{0}\dket{C_3}$ & $2\sqrt{6} \dket{1}\dket{j=2,m=0}$ & $4\dket{0}\dket{C_3}$ \\
$\dket{1}\dket{C_3}$ & $4\dket{0}\dket{1111}$ & $4\dket{1}\dket{C_3}$ \\
$\dket{0}\dket{C_1'}$ & $0$ & $-4 \dket{0}\dket{C_1'}$ \\
$\dket{1}\dket{C_1'}$ & $2\sqrt{2}\dket{0}\dket{D}$ & $-4 \dket{1}\dket{C_1'}$ \\
$\dket{0}\dket{C_3'}$ & $2\sqrt{2}\dket{1}\dket{D}$ & $-4 \dket{0}\dket{C_3'}$ \\
$\dket{1}\dket{C_3'}$ & $0$ & $-4 \dket{1}\dket{C_3'}$ \\
$\dket{0}\dket{D}$ & $2\sqrt{2}\dket{1}\dket{C_1'}$ & $0$ \\
$\dket{1}\dket{D}$ & $2\sqrt{2}\dket{0}\dket{C_3'}$ & $0$ \\
\hline
\end{tabular}
\end{center}
\caption{This table displays the action of the star and ring Hamiltonian on some important states} \label{tbl:sraction}
\end{table}

Having set out the relevant states we now review the state vectors
for energy levels I and II at the extremes of the pure ring
($c=0$) and pure star ($c=1$).  The eigenstates for the pure $XX$
ring for $N=4$ are given by Wang in \cite{XXRings}.  Note that in
our model there is a central spin which is uncoupled from the
outer spins for $c=0$.  Consequently it doubles the degeneracy of
all the ring eigenstates.  Energy level I is the ground state for
the pure ring.  It has energy $-4{\cal J}\sqrt{2}$ and it is a
mixture of
\begin{eqnarray}
\dket{0}\rt\left( \dket{A} - \dket{B} \right) \nonumber\\
\dket{1}\rt\left( \dket{A} - \dket{B} \right)
\label{eqn:srcpurering}
\end{eqnarray}
 Energy level II for the
pure ring is a mixture of
\[ \ba{l}
\dket{0}\dket{C_3'} \\
\dket{1}\dket{C_1'}
\ea \]
In the case of the pure star the energy eigenstates are given in
\cite{starspin}.
Energy level I for the pure star is a mixture of
\begin{eqnarray}
\rt \left( \dket{0}\dket{C_3} - \dket{1}\frac{1}{\sqrt{6}} \left( \sqrt{2} \dket{A} + 2\dket{B} \right) \right) \nonumber \\
\rt \left( \dket{0}\frac{1}{\sqrt{6}} \left( \sqrt{2} \dket{A} +
2\dket{B} \right) - \dket{1}\dket{C_1} \right)
\label{eqn:srcpurestar} \end{eqnarray}
 Energy level II for the pure star is a mixture of
\[ \ba{l}
\rt \left( \dket{0}\dket{C_3'} - \dket{1}\dket{D} \right) \\
\rt \left( \dket{0}\dket{D} - \dket{1}\dket{C_1'} \right)
\ea \]
A few remarks on degeneracy: From Table \ref{tbl:sraction} it is apparent that some states have been omitted here.  For example, $\dket{1}\dket{C_3'}$ and $\dket{0}\dket{C_1'}$ are also eigenstates for the pure ring at this point of energy level II.
However, from the energy level diagram in \cfig{fig:energies} it can be seen that for $c \neq 0$ the energy level splits and some states move up towards $E=0$ as $c$ approaches 1.  Therefore, in the mixtures above we have only included states that feature throughout all of energy level II.  Similarly, at $c=1$ (the star end) there are additional states present in energy level II due to degeneracy in $j$.  However these additional states diverge from energy level II for $c<1$ and therefore they have also been omitted above.

\begin{figure}
\begin{center}
\includegraphics[width=0.75\textwidth,clip]{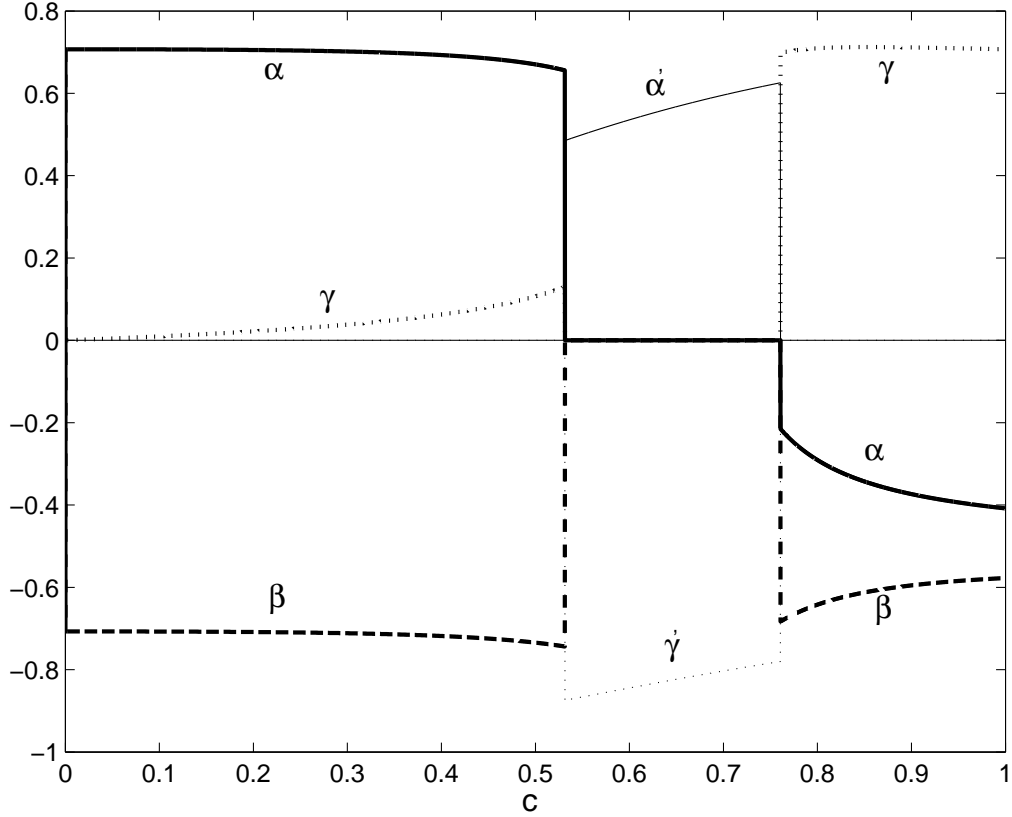}
\end{center}
\caption{This figure plots the coefficients
$\alpha,\beta,\gamma,\alpha^{'}$ and $\gamma^{'}$ involved in the
expressions for the ground states of $N=4$ for various values of
the parameter $c$.} \label{fig:regions}
\end{figure}

Given these extremes, we can form an impression of how the ring
state must mutate into the star state as $c$ changes from 0 to 1.
Our next step is to write down a general expression for the state
of each energy level which can cover a range of values of $c$.  To
do this, we split the graph into three regions - a `ring' region,
a `star' region, and an `intermediate' region.  These regions are
separated by the energy level crossings, causing discontinuities
in the entanglement. \cfig{fig:regions} illustrates these regions.

\subsection{The ring and star regions}

In the ground state the ring and star regions correspond to energy
level I. It is straightforward to write down a general expression
for the two degenerate eigenstates in energy level I:
\begin{equation} \ba{l}
\gamma \dket{0}\dket{C_3} + \dket{1}\left( \alpha \dket{A} + \beta
\dket{B}
\right) \\
\dket{0} \left( \alpha\dket{A} + \beta\dket{B} \right) +
\gamma\dket{1}\dket{C_1} \ea \end{equation} where $\alpha$,
$\beta$ and $\gamma$ are all functions of $c$.  From
\ceqn{eqn:srcpurering} and \ceqn{eqn:srcpurestar}  we see that at
$c=0$ (the ring), $\alpha=\rt$, $\beta=-\rt$ and $\gamma=0$.  At
$c=1$ (the star), $\alpha=-\sqrt{\frac{1}{6}}$,
$\beta=-\sqrt{\frac{2}{6}}$ and $\gamma=\rt$. The coefficients
$\alpha$, $\beta$ and $\gamma$ have been calculated numerically as
functions of $c$ and are plotted in \cfig{fig:regions}.

The reduced density matrix for nearest neighbors that arises when
we take an equal mixture of the above two states is
\[ \ba{l}
\frac{\beta^2}{4} \left( \dket{00}\dbra{00} + \dket{11}\dbra{11}
\right) + \dket{\psi^+(\frac{\alpha}{\sqrt{2}}, \frac{\beta}{2})}
\dbra{\psi^+(\frac{\alpha}{\sqrt{2}}, \frac{\beta}{2})} +
\dket{\psi^+(\frac{\beta}{2}, \frac{\alpha}{\sqrt{2}})}
\dbra{\psi^+(\frac{\beta}{2}, \frac{\alpha}{\sqrt{2}})} + \\
\frac{\gamma^2}{4} \left( 2 \dket{\Psi^+}\dbra{\Psi^+} +
\dket{00}\dbra{00} + \dket{11}\dbra{11} \right) \ea \] where we
have defined $\dket{\psi^+(u,v)} = \left( u \dket{01} + v
\dket{10} \right)$.  This gives a concurrence
\[ C = 2 \max \left\{ 0, \left| \frac{\gamma^2}{4} +
\frac{\alpha\beta}{\sqrt{2}} \right| - \frac{1}{4} \left( \gamma^2
+ \beta^2 \right) \right\} \]
Note that in the star region, $\alpha$ and $\beta$ are of the same
sign, which essentially makes the concurrence proportional to
$\alpha-\beta/2\sqrt{2}$, and thus it decreases with decreasing
$c$ because the $\alpha/\beta$ ratio decreases. It vanishes when
the ratio falls below $1/2\sqrt{2}$. In the ring region, $\alpha$
and $\beta$ are of opposite sign and $|\alpha\beta|>\gamma^2$,
which gives concurrence as
$-\frac{\beta}{\sqrt{2}}(\alpha+\beta/2\sqrt{2})-\gamma^2/2$. In
the ring region $\alpha+\beta/2\sqrt{2}$ is positive and the
concurrence decreases with increasing $c$ because $\gamma^2$
increases.

   For next-to-nearest neighbours, the reduced density matrix is
\[ \ba{l}
\frac{\alpha^2}{2} \left( \dket{00}\dbra{00} + \dket{11}\dbra{11}
\right) +
\beta^2\dket{\Psi^+}\dbra{\Psi^+} + \\
\frac{\gamma^2}{4} \left( 2 \dket{\Psi^+}\dbra{\Psi^+} +
\dket{00}\dbra{00} + \dket{11}\dbra{11} \right) \ea \] and this
gives a concurrence
\[ C = 2 \max \left\{ 0, \frac{1}{2} \left( \beta^2 - \alpha^2 \right)
\right\} \]
This expression for the concurrence neatly explains how the next
to nearest neighbor concurrence for $N=4$ varies with $c$- in the
ring region $|\alpha| \approx |\beta|$ and the next to nearest
neighbor concurrence is low, whereas in the star region $|\beta|
> |\alpha|$ and thus the concurrence is higher.

\subsection{The intermediate region}

Finally we consider the intermediate region, in which energy level
II is the ground state.  Given the two extremes at the star and
ring ends for energy level II above, we postulate the state is a
mixture of
\begin{eqnarray}
\gamma' \dket{0}\dket{C_3'} + \alpha'\dket{1}\dket{D}
\label{interstate1}\\
\alpha' \dket{0}\dket{D} - \gamma'\dket{1}\dket{C_1'}
\label{interstate2}
\end{eqnarray}
 The nearest neighbour reduced density matrix is
given by
\[
\frac{\gamma'^2}{4} \left( \dket{11}\dbra{11} + \dket{00}\dbra{00}
+ 2 \dket{\Psi^-}\dbra{\Psi^-} \right) + \frac{\alpha'^2}{2}
\left( \dket{01}\dbra{01} + \dket{10}\dbra{10} \right)
\]
which gives concurrence $C=0$. The next-to-nearest reduced density
matrix is given by
\[
\frac{\gamma'^2}{4} \left( \dket{11}\dbra{11} + \dket{00}\dbra{00}
+ 2 \dket{\Psi^+}\dbra{\Psi^+} \right) + \frac{\alpha'^2}{2}
\left( \dket{00}\dbra{00} + \dket{11}\dbra{11} \right)
\]
which also gives concurrence $C=0$.  Hence we have shown that this
energy level always has zero entanglement for both nearest and
next to nearest neighbors.

In summary, we have analysed the form of the state vectors of the energy levels I and II at the extremes of the star and ring models.  We have then interpolated the states to give general forms which cover the range $0<c<1$.  Using these general forms for the states we have obtained analytic formulae for the concurrence which match the numerical results.

\subsection{The production of GHZ and other multiparticle entangled states}
   As a final thought in this section, we note that, the
intermediate
   region, in which concurrence turns out to be zero, is not
   entirely uninteresting. In fact, it can even be regarded as the
   most interesting region of the model because it allows the production of a four particle GHZ (Greenberger-Horne-Zeilinger)
   state. Suppose we apply a magnetic field to separate the degenerate states of Eqs.(\ref{interstate1}) and (\ref{interstate2})
   in the intermediate region and
   follow it up by a measurement of the
central spin.  Then the state of the outer spins can be projected
with a probability of $|\alpha^{'}|^2$ (which is reasonably high,
namely $0.25$ to $0.36$ in the intermediate region) onto the state
$\dket{D} = \rt \left( \dket{0101} - \dket{1010} \right)$. This is
a very interesting state because, for all bipartite partitions of
the system, the state is maximally entangled.  This state is an
example of a four particle GHZ state. To our knowledge, there does
not yet exist any simple scheme for producing a GHZ state from the
ground state of a system of interacting spins. In our case, of
course, both the application of a magnetic field and the
measurement on the central spin are crucial. However, we can
regard the GHZ state as a ``simple derivative" of our ground state
in the intermediate region. Even when the production of the GHZ
state is unsuccessful, the state of the outer spins is projected
to yet another type of interesting multiparticle entangled state
namely $\dket{C_3'}$ or $\dket{C_1'}$. This state has the property
that the concurrence between any two spins is $2/N$, which is the
{\em maximum} possible entanglement in a collection of $N$ spins
in which all pairs of spins are equally entangled \cite{koashi}(in
the present case, $N=4$).

    The analysis also shows the {\em robustness} of the process of multiparticle entangled state
    production from ground states of spin stars \cite{starspin}. Any star geometry with a sufficiently
large number of outer spins placed in a ring will have them
physically close and will thereby add unwanted ring interactions
(interactions of an outer spin with its neighbors). Our
calculations here show that throughout the star region (as long as
the ring interactions are not too strong), we can produce the
states $\dket{C_3}$ or $\dket{C_1}$ which have the property that
the concurrence between any two spins is $2/N$. This happens with
a probability $|\gamma|^2$ (about $0.49$) in the star region when
the degeneracy of the ground states is lifted by a magnetic field
and a measurement is performed on the central spin. Thus in the
same way as described for the pure star in Ref.\cite{starspin},
multi-particle entangled states can be produced in a star polluted
with some degree of ring interaction.


\section{Ground states for general $N$ \label{sec:generalbonds}}

In this section we attempt to give a general explanation of the form of the ground state that applies to general values of $N$.  We start from the observation that, if two spins interact with each other via the $XX$ interaction then the ground state is the singlet state.  That is, the energy is minimised when the two spins are antiparallel.  As a general hypothesis then, we suppose that, in the combined star and ring model ground state, the $XX$ interaction generally tries to put adjacent spins into a singlet state.  Below we apply this hypothesis to even and odd $N$ separately and consider the effect that this would have on the ground state.

\begin{figure}[ht]
\begin{center}
\includegraphics[height=0.5in, clip]{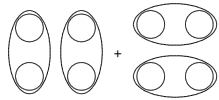}
\end{center}
\caption{A speculated state for $N=4$.} \label{fig:4pic}
\end{figure}

\begin{figure}[ht]
\begin{center}
\includegraphics[height=0.5in, clip]{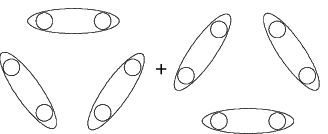}
\end{center}
\caption{A speculated state for $N=6$.} \label{fig:6pic}
\end{figure}

\begin{figure}[ht]
\begin{center}
\includegraphics[width=3in, clip]{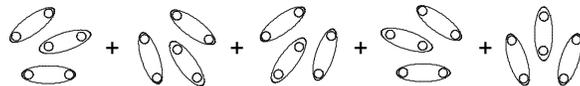}
\end{center}
\caption{A speculated state for $N=5$.} \label{fig:5pic}
\end{figure}

Firstly we consider the case $N=4$.  If we expect the $XX$
interaction to form singlets, then when in a ring we might expect
the superposition of singlet states depicted graphically in
\cfig{fig:4pic}. Here and henceforth (in the figures
\ref{fig:4pic}-\ref{fig:5pic}) the sign $+$ in the superposition
should be quite generally interpreted to mean superposition with a
general phase. We have always varied these phases to optimize the
overlap of our test states (those in figures
\ref{fig:4pic}-\ref{fig:5pic}) with the actual ground states.

To determine how close this superposition is to to the ground
state of the star-ring combination the fidelity between the ground
state and this superposition was calculated and is plotted in the
top left graph in \cfig{fig:ovs} where the fidelity is marked by
$O_p$. As can be seen, the fidelity is very close to 1 on the
ring-side, giving credence to the intuitive state.  As $c$
increases, the central spin is brought in to interact with the
other spins.  There is consequently an odd number of spins and the
system cannot form singlets.  The system is said to be
`frustrated'.  We would expect this to decrease the
nearest-neighbour entanglement and this is exactly what happens
while energy level I is the ground state - the nearest neighbour
entanglement decreases as $c$ increases.


It is natural to inquire whether a similar pattern holds for other
even $N$.  The bottom-right plot in \cfig{fig:ovs} compares the
superposition for $N=6$ depicted in \cfig{fig:6pic} with the
ground state for $N=6$ as $c$ varies from $0$ to $1$.
The fidelity of the superposition in \cfig{fig:6pic} isn't quite as high as it was for $N=4$ but is nevertheless quite high at approximately $0.9$.  This reinforces the idea that increasing $c$ causes the system to become increasingly frustrated with the consequence that the nearest neighbour entanglement decreases.

Next we apply the same hypothesis to the case of odd $N$, taking $N=5$ as an illustrative example.  In this case, for $c=0$ i.e.\ pure ring, one of the spins is unpaired, or frustrated.  As $c$ is increased, allowing interactions through the central spin, the frustrated spin may pair up with the central qubit is some manner and become less frustrated.  In \cfig{fig:5pic} we depict a superposition of states which represents this idea that the central spin pairs up with an `outer' qubit, taking account of the rotational symmetry of the model.

This model in \cfig{fig:5pic} could explain the variation of the entanglement for odd $N$ with $c$ - as $c$ increases the central spin somehow allows all the spins to pair up and form singlets, relieving the frustration.  If nearest-neighbour spins are `singlet-like' then are more likely to be entangled and thus the nearest-neighbour entanglement rises, as is indeed the case.  To provide some confirmation of this hypothesis the ground state of the star-ring combination can be compared to the superposition depicted in \cfig{fig:5pic} by calculating the fidelity between them.  This fidelity, marked by $O_p$, is plotted in the top right plot in \cfig{fig:ovs}.  This graph shows that this speculated state is indeed very close the ground state, and in fact, near $c=0.7$, \emph{is} the ground state.  Interestingly, the point at which it exactly becomes the ground state is also the point at which a different energy level takes over, leading to the sudden drop in nearest-neighbour entanglement.


We have shown that the hypothesis that the $XX$ interaction encourages spins to form singlets not only provides an explanation for the behaviour of the concurrence, but also gives a close approximation to the ground states as evidences by having a fidelity with the ground state approaching 1.


\subsection{Discussion}

In this section we first highlight some of the salient
observations in the paper:

\begin{itemize}
\item In section \ref{sec:elevelcrossings} it was pointed out that discontinuities in concurrence occured when the ground state changed.

\item The behavior of the concurrence as $c$ varied in the regions
were concurrence varies smoothly is \emph{not} due to the
crossings of energy levels. The example of $N=6$ in section
\ref{sec:elevelcrossings} showed that even though the same energy
level was the ground state for for $0.4<c<1$ the patterns observed
in the concurrence, such as a peak in the next to
nearest-neighbour concurrence for $c$ slightly greater than $0.7$
still occur.

\item Section \ref{sec:changingeig} indicated that the ground state stays `ring-like' and `star-like' for some time near $c=0$ and $c=1$ respectively.

\item Section \ref{sec:generalbonds} gave a reasonable explanation
for why the concurrence increases as $c$ was increased from $c=0$
for $N$ odd and decreased for $N$ even based on frustration.

\item In the intermediate region, at least for $N=4$, a GHZ state
can be produced as a simple derivative of the ground state.

\item Some interesting multiparticle entangled states which can be
produced in a star system of spins can also be produced even when
a significant proportion of ring interaction is present on top of
the star interaction.

\end{itemize}

    In Section \ref{sec:generalbonds} we have already given an explanation for the apparently counter-intuitive
the rise of the nearest neighbor entanglement on increasing $c$
from the ring side for odd $N$. We now attempt to provide an
explanation for the rest of the behavior of the entanglement in
the system that we have numerically observed. This includes the
apparently counter-intuitive rise of the next to nearest neighbor
entanglement as $c$ is decreased in the star region and the
vanishing of both nearest and next to nearest neighbor
entanglement around $c=0.7$. We will provide the explanation only
in the case of $N=4$, and assume that an analogous argument holds
for other $N$. In the case of $N=4$, we have already provided a
mathematical explanation of the behavior of entanglement (in
section\ref{sec:expln4}) by accepting certain numerically observed
patterns for the behavior of $\alpha,\beta$ and $\gamma$. It is
these patterns that we will now explain by taking for granted the
numerically observed facts that the coefficients $\alpha,\beta$
and $\gamma$ always remain real and they smoothly change to
interpolate between the ground state of the star and the ring as
$c$ varies.

   In the ground state at the star ($c=1$) end, the outer spins are in a mixture of fully symmetric
states. This is because the star Hamiltonian converts states
$\dket{1} \dket{C_1}$ to $\dket{0}\frac{1}{\sqrt{6}} \left(
\sqrt{2} \dket{A} + 2\dket{B} \right)$ (and $\dket{0} \dket{C_3}$
to $\dket{1}\frac{1}{\sqrt{6}} \left( \sqrt{2} \dket{A} +
2\dket{B} \right)$) and vice versa. The star interaction thus
tends to {\em symmetrize} the state of the outer spins ({\em
i.e.,} impose the same sign on $\alpha$ and $\beta$) and impose an
opposite sign to $\gamma$ with respect to $\alpha$ and $\beta$ in
the ground state. On the other hand, the ring interaction converts
two subparts of a symmetric state (namely $\dket{A}$ and
$\dket{B}$) to each other, and thus tends to impose a sign
difference between these states in the ground state. Indeed the
ground state at the ring end ($c=0$) consists of the state
$\frac{1}{\sqrt{2}}(\dket{A}-\dket{B})$ irrespective of the state
of the central spin. To {\em smoothly interpolate} between the
ground states of the star and the ring by a single ground state,
then, $\alpha$ and $\beta$ have to change from being of the same
sign to being of opposite signs as one proceeds from the star end
to the ring end. In order to do this, one of them (either $\alpha$
or $\beta$) has to retain its sign and thereby remain opposite in
sign to $\gamma$, while the other has to go through zero and
reverse its sign. In the star region, the effect of the star
interaction is strong, and in this region the energy is lowest if
$\beta$, rather than $\alpha$, is maintained to be opposite in
sign from $\gamma$. This is simply because the energy of a state
of the type $\gamma \dket{0}\dket{C_3} + \dket{1}\left( \alpha
\dket{A} + \beta \dket{B} \right)$ with positive $\gamma$ and
negative $\alpha$ and $\beta$ is lower for a larger proportion of
$\dket{B}$ rather than for a larger proportion of $\dket{A}$. The
value of $\alpha$, which should reverse sign, will thus move
towards zero as $c$ is decreased from the star end. This decreases
the proportion of $\dket{A}$ in the ground state. As $\dket{A}$
has nearest neighbors in opposite states, it constructively
contributes to the entanglement of nearest neighbors. If it
decreases, so does the nearest neighbor entanglement. As far as
next to nearest neighbor state is concerned, though, $\dket{A}$
contributes only $|00\rangle$ or $|11\rangle$ to the state. From
the generic expression (Eq.(\ref{concexpression})) for two spin
reduced density matrices for this system, we know that
entanglement can only stem from the presence of states
$|01\rangle$ or $|10\rangle$. So $\dket{A}$ does not contribute
positively to next to nearest neighbor entanglement. On the other
hand, it contributes a fraction of unentangled states
$|00\rangle\langle 00|$ and $|11\rangle\langle 11|$ to the mixed
state of the next to nearest neighbor qubits, which reduces the
entanglement in the state. Thus when the fraction of $\dket{A}$
decreases, the part of the state which contributes to next to
nearest neighbor entanglement increases due to normalization,
thereby increasing this entangelement. This explains the
apparently counterintuitive rise of the next to nearest neighbor
entanglement in the star region as $c$ decreases.

     The decrease in the proportion of $\dket{A}$ in the state,
however, increases the energy due to the ring part of the
interaction, as this part of the interaction ``prefers" ({\em
i.e.,} lowers the energy of) states in which nearest neighbors are
oppositely aligned. The energy of the state thus continues to
increase as $c$ decreases. For $0.7\leq c \leq 1$, however, it
still continues to be the lowest energy state because of the
dominance of the star interaction in this region. Around $c\approx
0.7$ the buildup of energy due to decreasing proportion of
$\dket{A}$ is not sustainable, and a different pair of states
overtake as the ground state. These states, which signal the start
of the intermediate region, are of the form $\gamma'
\dket{0}\dket{C_3'} + \alpha'\dket{1}\dket{D}$ and $\alpha'
\dket{0}\dket{D} - \gamma'\dket{1}\dket{C_1'}$. This has a
significant proportion of $\dket{D}$, which, because of its
similar nature as $\dket{A}$, lowers the energy due to the ring
part of the interaction. The state of the outer spins
corresponding to this state is a mixture of a state $\dket{C_1'}$,
$\dket{C_3'}$ and $\dket{D}$. Both $\dket{D}$, and an equal
mixture of $\dket{C_1'}$ and $\dket{C_3'}$, individually have zero
nearest neighbor and next to nearest neighbor entanglement, which
explains the dropping of all entanglement to zero for a region
after $c\approx 0.7$ (the intermediate region).

   The state in the intermediate region still has a significant
absolute value of $\gamma'$, which plays a role in lowering the
energy due to the star part of the interaction. However, as we
approach the end of the intermediate region by decreasing $c$, the
star interaction becomes altogether less important, and then it is
more important to have a state of the form $\alpha \dket{A}
+\beta\dket{B}$ with $\alpha$ and $\beta$ of comparable absolute
values but opposite in sign to maximally lower the energy due to
the dominant ``ring" part of the interaction. At this value of
$c$, the energy level which smoothly interpolates between the
ground states of the star and the ring has assumed precisely such
a form (except for a very small extra fraction of $\gamma\dket{1}
\dket{C_1'}$ or $\gamma\dket{0}\dket{C_3'}$), and becomes the
ground state once again. This indicates the start of the ring
region when $c$ is decreased.

\section{Conclusions \label{sec:srcomboconc}}

This paper has demonstrated that the model in which the outer
spins can interact through a combination of star and ring type
interactions possesses a number of surprising features which make
it interesting to study.  In this paper we have shown that both
nearest neighbor (for odd $N$) and next to nearest neighbor (for
all $N$) entanglement in the ground state have their maxima for a
Hamiltonian which is {\em neither} a pure ring {\em nor} a pure
star in its interactions. We have drawn attention to the link
between dramatic changes in entanglement and the change in the
ground state due to crossing over of energy levels.  The case of
four outer spins was analyzed in detail and interpolated ground
states given for all values of $c$.  By hypothesizing a tendency
of the interaction to form singlet states we have found an
explanation for the behavior of the entanglement in the ring
region that applies to general values of $N$. We have found that
we can produce a GHZ state as a simple derivative of the ground
state for $N=4$. We have also shown that the multi-particle
entangled states producible from a pure star are also producible
from in a star system polluted with a significant proportion of
ring interaction.

We believe the concept of a combination of interactions by both
models is relevant because although it is unlikely to be a
naturally occuring structure, experimental implementations of
quantum computing, for example using quantum dots, may allow
artifical structures to be created where the topology is in fact
the combination we have been describing.  In that case our results
will be useful, especially is situations where there are untunable
(fixed) interactions.

There are a number of potential avenues for future working
stemming from the material described here.  It would be satisfying
to be able to expand the range of $N$ considered to try to spot
broader trends and patterns. Indeed to fully describe the model a
complete analytical solution would be desirable although this is
most likely very difficult to find. The concept of a network of
spins interacting through a combination of two different
topologies could perhaps be extended to other structures and
dimensions. Recently, it has been shown that spin systems can be
used for studying non-Markovian dynamics \cite{burgarth}, optimal
quantum cloning \cite{dechiara} (where a spin star can be used)
and quantum computation \cite{benjamin}. It would be interesting
to investigate the dynamical consequences of spins interacting in
a {\em combination} of star and ring geometries.

\section{Acknowledgements}
AH thanks UK EPSRC for financial support. Part of this work was
carried out when SB was a postdoctoral scholar (supported by the
NSF under Grant Number
   EIA-00860368) and AH was a visitor at the Institute for Quantum
Information, Caltech, where we thank the hospitality of John
Preskill. We thank Daniel Burgarth for a careful reading of the
manuscript and valuable comments.


\begin{thebibliography}{34}
\expandafter\ifx\csname natexlab\endcsname\relax\def\natexlab#1{#1}\fi
\expandafter\ifx\csname bibnamefont\endcsname\relax
  \def\bibnamefont#1{#1}\fi
\expandafter\ifx\csname bibfnamefont\endcsname\relax
  \def\bibfnamefont#1{#1}\fi
\expandafter\ifx\csname citenamefont\endcsname\relax
  \def\citenamefont#1{#1}\fi
\expandafter\ifx\csname url\endcsname\relax
  \def\url#1{\texttt{#1}}\fi
\expandafter\ifx\csname urlprefix\endcsname\relax\def\urlprefix{URL }\fi
\providecommand{\bibinfo}[2]{#2}
\providecommand{\eprint}[2][]{\url{#2}}

\bibitem[{\citenamefont{O'Connor and Wootters}(2001)}]{EntRings}
\bibinfo{author}{\bibfnamefont{K.~M.} \bibnamefont{O'Connor}} \bibnamefont{and}
  \bibinfo{author}{\bibfnamefont{W.~K.} \bibnamefont{Wootters}},
  \bibinfo{journal}{Phys. Rev. A} \textbf{\bibinfo{volume}{63}},
  \bibinfo{pages}{052302} (\bibinfo{year}{2001}).

\bibitem[{\citenamefont{Nielsen}(1998)}]{thermentA}
\bibinfo{author}{\bibfnamefont{M.~A.} \bibnamefont{Nielsen}},
\emph{\bibinfo{title}{Quantum Information Theory}}, Ph.D. thesis,
  \bibinfo{school}{University of New Mexico} (\bibinfo{year}{1998}),
  \eprint{quant-ph/0011036}.

\bibitem[{\citenamefont{Arnesen et~al.}(2001)\citenamefont{Arnesen, Bose, and
  Vedral}}]{ThermE-1D}
\bibinfo{author}{\bibfnamefont{M.~C.} \bibnamefont{Arnesen}},
  \bibinfo{author}{\bibfnamefont{S.}~\bibnamefont{Bose}}, \bibnamefont{and}
  \bibinfo{author}{\bibfnamefont{V.}~\bibnamefont{Vedral}},
  \bibinfo{journal}{Phys. Rev. Lett.} \textbf{\bibinfo{volume}{87}},
  \bibinfo{pages}{017901} (\bibinfo{year}{2001}).





\bibitem[{\citenamefont{Wang}(2002)}]{XXRings}
\bibinfo{author}{\bibfnamefont{X.}~\bibnamefont{Wang}}, \bibinfo{journal}{Phys.
  Rev. A} \textbf{\bibinfo{volume}{66}}, \bibinfo{pages}{034302}
  (\bibinfo{year}{2002}).



\bibitem[{\citenamefont{Wang}(2001{\natexlab{a}})}]{EntXY}
\bibinfo{author}{\bibfnamefont{X.}~\bibnamefont{Wang}}, \bibinfo{journal}{Phys.
  Rev. A} \textbf{\bibinfo{volume}{64}}, \bibinfo{pages}{012313}
  (\bibinfo{year}{2001}{\natexlab{a}}).

\bibitem[{\citenamefont{Wang}(2001{\natexlab{b}})}]{ThermXXZ}
\bibinfo{author}{\bibfnamefont{X.}~\bibnamefont{Wang}}, \bibinfo{journal}{Phys.
  Lett. A} \textbf{\bibinfo{volume}{281}}, \bibinfo{pages}{101}
  (\bibinfo{year}{2001}{\natexlab{b}}).

\bibitem[{\citenamefont{Gunlycke et~al.}(2001)\citenamefont{Gunlycke,
  Kendon, Vedral and Bose}}]{IsingVarB}
\bibinfo{author}{\bibfnamefont{D.}~\bibnamefont{Gunlycke}},
\bibinfo{author}{\bibfnamefont{V.~M.} \bibnamefont{Kendon}},
  \bibnamefont{and} \bibinfo{author}{\bibfnamefont{V.}~\bibnamefont{Vedral}},
  \bibinfo{author}{\bibfnamefont{S.}~\bibnamefont{Bose}},
  \bibinfo{journal}{Phys. Rev. A} \textbf{\bibinfo{volume}{64}},
  \bibinfo{pages}{042302} (\bibinfo{year}{2001}).



\bibitem[{\citenamefont{Wang et~al.}(2002)\citenamefont{Wang, Fu, and
  Solomon}}]{wang01-3}
\bibinfo{author}{\bibfnamefont{X.}~\bibnamefont{Wang}},
  \bibinfo{author}{\bibfnamefont{H.}~\bibnamefont{Fu}}, \bibnamefont{and}
  \bibinfo{author}{\bibfnamefont{A.~I.} \bibnamefont{Solomon}},
  \bibinfo{journal}{J. Phys A: Math. Gen.} \textbf{\bibinfo{volume}{35}},
  \bibinfo{pages}{4293} (\bibinfo{year}{2002}).

\bibitem[{\citenamefont{Kamta and Starace}(2002)}]{kamta}
\bibinfo{author}{\bibfnamefont{G.~L.} \bibnamefont{Kamta}} \bibnamefont{and}
  \bibinfo{author}{\bibfnamefont{A.~F.} \bibnamefont{Starace}},
  \bibinfo{journal}{Phys. Rev. Lett.} \textbf{\bibinfo{volume}{88}},
  \bibinfo{pages}{107901} (\bibinfo{year}{2002}).

\bibitem[{\citenamefont{Osterloh et~al.}(2002)\citenamefont{Osterloh, Amico,
  Falci, and Fazio}}]{falci-1}
\bibinfo{author}{\bibfnamefont{A.}~\bibnamefont{Osterloh}},
  \bibinfo{author}{\bibfnamefont{L.}~\bibnamefont{Amico}},
  \bibinfo{author}{\bibfnamefont{G.}~\bibnamefont{Falci}}, \bibnamefont{and}
  \bibinfo{author}{\bibfnamefont{R.}~\bibnamefont{Fazio}},
  \bibinfo{journal}{Nature} \textbf{\bibinfo{volume}{416}},
  \bibinfo{pages}{608} (\bibinfo{year}{2002}).

\bibitem[{\citenamefont{Osborne and Nielsen}(2002)}]{falci-2}
\bibinfo{author}{\bibfnamefont{T.~J.} \bibnamefont{Osborne}} \bibnamefont{and}
  \bibinfo{author}{\bibfnamefont{M.~A.} \bibnamefont{Nielsen}},
  \bibinfo{journal}{Phys. Rev. A} \textbf{\bibinfo{volume}{66}},
  \bibinfo{pages}{032110} (\bibinfo{year}{2002}).

\bibitem[{\citenamefont{Bose and Chattopadhyay}(2002)}]{ibose}
\bibinfo{author}{\bibfnamefont{I.}~\bibnamefont{Bose}} \bibnamefont{and}
  \bibinfo{author}{\bibfnamefont{E.}~\bibnamefont{Chattopadhyay}},
  \bibinfo{journal}{Phys. Rev. A} \textbf{\bibinfo{volume}{66}},
  \bibinfo{pages}{062320} (\bibinfo{year}{2002}).

\bibitem[{\citenamefont{Lakshminarayan and Subrahmanyam}(2003)}]{arul}
\bibinfo{author}{\bibfnamefont{A.}~\bibnamefont{Lakshminarayan}}
  \bibnamefont{and}
  \bibinfo{author}{\bibfnamefont{V.}~\bibnamefont{Subrahmanyam}},
  \bibinfo{journal}{Phys. Rev. A} \textbf{\bibinfo{volume}{67}},
  \bibinfo{pages}{052304} (\bibinfo{year}{2003}).

\bibitem[{\citenamefont{Audenaert et~al.}(2002)\citenamefont{Audenaert, Eisert,
  Plenio, and Werner}}]{plenio}
\bibinfo{author}{\bibfnamefont{K.}~\bibnamefont{Audenaert}},
  \bibinfo{author}{\bibfnamefont{J.}~\bibnamefont{Eisert}},
  \bibinfo{author}{\bibfnamefont{M.~B.} \bibnamefont{Plenio}},
  \bibnamefont{and} \bibinfo{author}{\bibfnamefont{R.~F.}
  \bibnamefont{Werner}}, \bibinfo{journal}{Phys. Rev. A}
  \textbf{\bibinfo{volume}{66}}, \bibinfo{pages}{042327}
  (\bibinfo{year}{2002}).

\bibitem[{\citenamefont{Vidal et~al.}(2003)\citenamefont{Vidal, Latorre, Rico,
  and Kitaev}}]{guifre}
\bibinfo{author}{\bibfnamefont{G.}~\bibnamefont{Vidal}},
  \bibinfo{author}{\bibfnamefont{J.~I.} \bibnamefont{Latorre}},
  \bibinfo{author}{\bibfnamefont{E.}~\bibnamefont{Rico}}, \bibnamefont{and}
  \bibinfo{author}{\bibfnamefont{A.}~\bibnamefont{Kitaev}},
  \bibinfo{journal}{Phys. Rev. Lett.} \textbf{\bibinfo{volume}{90}},
  \bibinfo{pages}{227902} (\bibinfo{year}{2003}).

\bibitem{vladimir}
V. E. Korepin, Phys. Rev. Lett. {\bf 92}, 096402 (2004).



\bibitem[{\citenamefont{Wang and Zanardi}(2002)}]{wangxx-2}
\bibinfo{author}{\bibfnamefont{X.}~\bibnamefont{Wang}} \bibnamefont{and}
  \bibinfo{author}{\bibfnamefont{P.}~\bibnamefont{Zanardi}},
  \bibinfo{journal}{Phys. Lett. A} \textbf{\bibinfo{volume}{301}},
  \bibinfo{pages}{1} (\bibinfo{year}{2002}).

\bibitem[{\citenamefont{Ghosh et~al.}(2003)\citenamefont{Ghosh, Rosenbaum,
  Aeppli, and Coppersmith}}]{aeppli-vedral-1}
\bibinfo{author}{\bibfnamefont{S.}~\bibnamefont{Ghosh}},
  \bibinfo{author}{\bibfnamefont{T.~F.} \bibnamefont{Rosenbaum}},
  \bibinfo{author}{\bibfnamefont{G.}~\bibnamefont{Aeppli}}, \bibnamefont{and}
  \bibinfo{author}{\bibfnamefont{S.~N.} \bibnamefont{Coppersmith}},
  \bibinfo{journal}{Nature} \textbf{\bibinfo{volume}{425}}, \bibinfo{pages}{48}
  (\bibinfo{year}{2003}).

\bibitem[{\citenamefont{Vedral}(2003)}]{aeppli-vedral-2}
\bibinfo{author}{\bibfnamefont{V.}~\bibnamefont{Vedral}},
  \bibinfo{journal}{Nature} \textbf{\bibinfo{volume}{425}}, \bibinfo{pages}{28}
  (\bibinfo{year}{2003}).

\bibitem[{\citenamefont{Vedral}()}]{vedral2}
\bibinfo{author}{\bibfnamefont{V.}~\bibnamefont{Vedral}},
\bibinfo{journal}{New. J. Phys.} \textbf{\bibinfo{volume}{6}}, \bibinfo{pages}{22}
(\bibinfo{year}{2004}).

\bibitem[{\citenamefont{Sachdev}(1999)}]{sachdev}
\bibinfo{author}{\bibfnamefont{S.}~\bibnamefont{Sachdev}},
  \emph{\bibinfo{title}{Quantum Phase Transitions}}
  (\bibinfo{publisher}{Cambridge University Press, Cambridge, U.K.},
  \bibinfo{year}{1999}).

\bibitem[{\citenamefont{Hutton and Bose}()}]{starspin}
\bibinfo{author}{\bibfnamefont{A.}~\bibnamefont{Hutton}} \bibnamefont{and}
  \bibinfo{author}{\bibfnamefont{S.}~\bibnamefont{Bose}},
   \bibinfo{journal}{Phys. Rev. A} \textbf{\bibinfo{volume}{69}},
 \bibinfo{pages}{042312} (\bibinfo{year}{2004}).


\bibitem[{\citenamefont{Imamo{\={g}}lu
  et~al.}(1999)\citenamefont{Imamo{\={g}}lu, Awschalom, Burkard, DiVincenzo,
  Loss, Sherwin, and Small}}]{QD+CavQED}
\bibinfo{author}{\bibfnamefont{A.}~\bibnamefont{Imamo{\={g}}lu}},
  \bibinfo{author}{\bibfnamefont{D.~D.} \bibnamefont{Awschalom}},
  \bibinfo{author}{\bibfnamefont{G.}~\bibnamefont{Burkard}},
  \bibinfo{author}{\bibfnamefont{D.~P.} \bibnamefont{DiVincenzo}},
  \bibinfo{author}{\bibfnamefont{D.}~\bibnamefont{Loss}},
  \bibinfo{author}{\bibfnamefont{M.}~\bibnamefont{Sherwin}}, \bibnamefont{and}
  \bibinfo{author}{\bibfnamefont{A.}~\bibnamefont{Small}},
  \bibinfo{journal}{Phys. Rev. Lett.} \textbf{\bibinfo{volume}{83}},
  \bibinfo{pages}{4204} (\bibinfo{year}{1999}).

\bibitem[{\citenamefont{Zheng and Guo}(2000)}]{ImplCavQED}
\bibinfo{author}{\bibfnamefont{S.~B.} \bibnamefont{Zheng}} \bibnamefont{and}
  \bibinfo{author}{\bibfnamefont{G.~C.} \bibnamefont{Guo}},
  \bibinfo{journal}{Phys. Rev. Lett.} \textbf{\bibinfo{volume}{85}},
  \bibinfo{pages}{2392} (\bibinfo{year}{2000}).

\bibitem[{\citenamefont{Cirac and Zoller}(2000)}]{ImplMTraps}
\bibinfo{author}{\bibfnamefont{J.~I.} \bibnamefont{Cirac}} \bibnamefont{and}
  \bibinfo{author}{\bibfnamefont{P.}~\bibnamefont{Zoller}},
  \bibinfo{journal}{Nature} \textbf{\bibinfo{volume}{404}},
  \bibinfo{pages}{579} (\bibinfo{year}{2000}).

\bibitem[{\citenamefont{Kane}(1998)}]{ImplSiNuc}
\bibinfo{author}{\bibfnamefont{B.~E.} \bibnamefont{Kane}},
  \bibinfo{journal}{Nature} \textbf{\bibinfo{volume}{393}},
  \bibinfo{pages}{133} (\bibinfo{year}{1998}).

\bibitem[{\citenamefont{Makhlin et~al.}(1999)\citenamefont{Makhlin, Schon, and
  Shnirman}}]{JJunctQubits}
\bibinfo{author}{\bibfnamefont{Y.}~\bibnamefont{Makhlin}},
  \bibinfo{author}{\bibfnamefont{G.}~\bibnamefont{Schon}}, \bibnamefont{and}
  \bibinfo{author}{\bibfnamefont{A.}~\bibnamefont{Shnirman}},
  \bibinfo{journal}{Nature} \textbf{\bibinfo{volume}{398}},
  \bibinfo{pages}{305} (\bibinfo{year}{1999}).

\bibitem[{\citenamefont{Benjamin and Bose}()}]{benjamin2}
\bibinfo{author}{\bibfnamefont{S.~C.} \bibnamefont{Benjamin}} \bibnamefont{and}
  \bibinfo{author}{\bibfnamefont{S.}~\bibnamefont{Bose}},
  \emph{\bibinfo{title}{Quantum computing in arrays coupled by 'always on'
  interactions}}, \eprint{quant-ph/0401071} (To appear in Phys. Rev. A).

\bibitem{dawson}
C. M. Dawson and M. A. Nielsen, Phys. Rev. A {\bf 69}, 052316
(2004).
\bibitem{ghz}
D. M. Greenberger, M. A. Horne, and A. Zeilinger, in {\em Bell's
Theorem, Quantum Theory, and Conceptions of the universe}, ed. M.
Kafatos (Kluwer, Dordrecht, 1989).


  \bibitem[{\citenamefont{Burgarth}()\citenamefont{Burgarth}}]{burgarth}
\bibinfo{author}{\bibfnamefont{H.-P.} \bibnamefont{Breuer}},
  \bibinfo{author}{\bibfnamefont{D.}~\bibnamefont{Burgarth}}
   \bibnamefont{and}
   \bibinfo{author}{\bibfnamefont{F.}~\bibnamefont{Petruccione}},
\bibinfo{journal}{Phys. Rev. B} \textbf{\bibinfo{volume}{70}},
 \bibinfo{pages}{045323} (\bibinfo{year}{2004}).

  \bibitem[{\citenamefont{Chiara et~al.}()\citenamefont{Chiara, Fazio,
  Macchiavello, Montangero, and Palma}}]{dechiara}
\bibinfo{author}{\bibfnamefont{G.~D.} \bibnamefont{Chiara}},
  \bibinfo{author}{\bibfnamefont{R.}~\bibnamefont{Fazio}},
  \bibinfo{author}{\bibfnamefont{C.}~\bibnamefont{Macchiavello}},
  \bibinfo{author}{\bibfnamefont{S.}~\bibnamefont{Montangero}},
  \bibnamefont{and} \bibinfo{author}{\bibfnamefont{G.~M.} \bibnamefont{Palma}},
  \emph{\bibinfo{title}{Quantum cloning in spin networks}},
  \eprint{quant-ph/0402071}.

\bibitem{benjamin3}
S. C. Benjamin, New J. Phys. {\bf 6}, 61 (2004).


\bibitem[{\citenamefont{Bennett et~al.}(1996)\citenamefont{Bennett, DiVincenzo,
  Smolin, and Wootters}}]{BigMixed}
\bibinfo{author}{\bibfnamefont{C.~H.} \bibnamefont{Bennett}},
  \bibinfo{author}{\bibfnamefont{D.~P.} \bibnamefont{DiVincenzo}},
  \bibinfo{author}{\bibfnamefont{J.~A.} \bibnamefont{Smolin}},
  \bibnamefont{and} \bibinfo{author}{\bibfnamefont{W.~K.}
  \bibnamefont{Wootters}}, \bibinfo{journal}{Phys. Rev. A}
  \textbf{\bibinfo{volume}{54}}, \bibinfo{pages}{3824} (\bibinfo{year}{1996}).

\bibitem[{\citenamefont{Wootters}(1998)}]{EntForm}
\bibinfo{author}{\bibfnamefont{W.~K.} \bibnamefont{Wootters}},
  \bibinfo{journal}{Phys. Rev. Lett.} \textbf{\bibinfo{volume}{80}},
  \bibinfo{pages}{2245} (\bibinfo{year}{1998}).

\bibitem[{\citenamefont{Hill and Wootters}(1997)}]{EntForm0}
\bibinfo{author}{\bibfnamefont{S.}~\bibnamefont{Hill}} \bibnamefont{and}
  \bibinfo{author}{\bibfnamefont{W.~K.} \bibnamefont{Wootters}},
  \bibinfo{journal}{Phys. Rev. Lett.} \textbf{\bibinfo{volume}{78}},
  \bibinfo{pages}{5022} (\bibinfo{year}{1997}).

\bibitem{Bethe}
H. A. Bethe, Z. Physik {\bf 71}, 205 (1931).

\bibitem[{\citenamefont{Verstraete et~al.}(2004)\citenamefont{Verstraete, Popp,
  and Cirac}}]{Localizable}
\bibinfo{author}{\bibfnamefont{F.}~\bibnamefont{Verstraete}},
  \bibinfo{author}{\bibfnamefont{M.}~\bibnamefont{Popp}}, \bibnamefont{and}
  \bibinfo{author}{\bibfnamefont{J.~I.} \bibnamefont{Cirac}},
  \bibinfo{journal}{Phys. Rev. Lett.} \textbf{\bibinfo{volume}{92}},
  \bibinfo{pages}{027901} (\bibinfo{year}{2004}).

\bibitem{jin}
B.-Q. Jin and V.E. Korepin,  Phys. Rev. A {\bf 69}, 062314 (2004).





\bibitem[{\citenamefont{Jozsa}(1994)}]{Fidelity}
\bibinfo{author}{\bibfnamefont{R.}~\bibnamefont{Jozsa}}, \bibinfo{journal}{J.
  Mod. Opt.} \textbf{\bibinfo{volume}{41}}, \bibinfo{pages}{2315}
  (\bibinfo{year}{1994}).

\bibitem{koashi}
M. Koashi, V. Buzek, and N. Imoto, Phys. Rev. A {\bf 62}, 050302
(2000).



\bibitem[{\citenamefont{Benjamin and Bose}(2003)}]{benjamin}
\bibinfo{author}{\bibfnamefont{S.~C.} \bibnamefont{Benjamin}} \bibnamefont{and}
  \bibinfo{author}{\bibfnamefont{S.}~\bibnamefont{Bose}},
  \bibinfo{journal}{Phys. Rev. Lett.} \textbf{\bibinfo{volume}{90}},
  \bibinfo{pages}{247901} (\bibinfo{year}{2003}).

\end{thebibliography}
\end{document}